\documentclass[a4paper,11pt]{article}
\usepackage{jcappub}

\usepackage{hyperref}
\usepackage{verbatim}
\usepackage{tabularx}
\usepackage{orcidlink}
\usepackage{amsmath}
\usepackage{appendix}
\usepackage {siunitx} 
\usepackage[utf8]{inputenc}
\usepackage{lineno}
\DeclareUnicodeCharacter{2212}{\textminus}
\DeclareRobustCommand{\ion}[2]{%
\relax\ifmmode
\ifx\testbx\f@series
{\mathbf{#1\,\mathsc{#2}}}\else
{\mathrm{#1\,\mathsc{#2}}}\fi
\else\textup{#1\,{\mdseries\textsc{#2}}}%
\fi}

\title{\boldmath High redshift LBGs from deep broadband imaging for future spectroscopic surveys}

\author[1]{{Vanina~Ruhlmann-Kleider}\orcidlink{0009-0000-6063-6121},}
\author[1]{{Christophe~Yèche}\orcidlink{0000-0001-5146-8533},}
\author[1]{{Christophe~Magneville},}
\author[1]{{Henri~Coquinot},}
\author[1]{{Eric~Armengaud}\orcidlink{0000-0001-7600-5148},}
\author[1,2]{{Nathalie~Palanque-Delabrouille}\orcidlink{0000-0003-3188-784X},}
\author[2]{{Anand~Raichoor}\orcidlink{0000-0001-5999-7923},}
\author[2]{{Jessica~Nicole~Aguilar},}
\author[3]{{Steven~Ahlen}\orcidlink{0000-0001-6098-7247},}
\author[4]{{St\'{e}phane~Arnouts},}
\author[5]{{David~Brooks},}
\author[2]{{Edmond~Chaussidon}\orcidlink{0000-0001-8996-4874},}
\author[2]{{Todd~Claybaugh},}
\author[6]{{Kyle~Dawson},}
\author[7]{{Axel~de~la~Macorra}\orcidlink{0000-0002-1769-1640},}
\author[8]{{Arjun~Dey}\orcidlink{0000-0002-4928-4003},}
\author[9]{{Biprateep~Dey}\orcidlink{0000-0002-5665-7912},}
\author[5]{{Peter~Doel},}
\author[10,11]{{Kevin~Fanning}\orcidlink{0000-0003-2371-3356},}
\author[2,12]{{Simone~Ferraro}\orcidlink{0000-0003-4992-7854},}
\author[13,14]{{Jaime~E.~Forero-Romero}\orcidlink{0000-0002-2890-3725},}
\author[2]{{Satya~Gontcho~A~Gontcho}\orcidlink{0000-0003-3142-233X},}
\author[15]{{Gaston~Gutierrez},}
\author[16]{{Stephen~Gwyn}\orcidlink{0000-0001-8221-8406},}
\author[17,18,19]{{Klaus~Honscheid},}
\author[8]{{Stephanie~Juneau},}
\author[20]{{Robert~Kehoe},}
\author[2]{{Theodore~Kisner}\orcidlink{0000-0003-3510-7134},}
\author[2]{{Anthony~Kremin}\orcidlink{0000-0001-6356-7424},}
\author[2]{{Andrew~Lambert},}
\author[2]{{Martin~Landriau}\orcidlink{0000-0003-1838-8528},}
\author[21]{{Laurent~Le~Guillou}\orcidlink{0000-0001-7178-8868},}
\author[2]{{Michael~E.~Levi}\orcidlink{0000-0003-1887-1018},}
\author[22,23]{{Marc~Manera}\orcidlink{0000-0003-4962-8934},}
\author[17,24,19]{{Paul~Martini}\orcidlink{0000-0002-4279-4182},}
\author[8]{{Aaron~Meisner}\orcidlink{0000-0002-1125-7384},}
\author[25,23]{{Ramon~Miquel},}
\author[26]{{John~Moustakas}\orcidlink{0000-0002-2733-4559},}
\author[27]{{Eva-Maria~Mueller},}
\author[7]{{Andrea~Muñoz-Gutiérrez},}
\author[9]{{Jeffrey~A.~Newman}\orcidlink{0000-0001-8684-2222},}
\author[28]{{Jundan~Nie}\orcidlink{0000-0001-6590-8122},}
\author[29,30]{{Gustavo~Niz}\orcidlink{0000-0002-1544-8946},}
\author[1]{{Constantin~Payerne},}
\author[31]{{Vincent~Picouet},}
\author[32,1,33]{{Corentin~Ravoux}\orcidlink{0000-0002-3500-6635},}
\author[34]{{Mehdi~Rezaie}\orcidlink{0000-0001-5589-7116},}
\author[35]{{Graziano~Rossi},}
\author[36]{{Eusebio~Sanchez}\orcidlink{0000-0002-9646-8198},}
\author[37]{{Marcin~Sawicki}\orcidlink{0000-0002-7712-7857},}
\author[38]{{Edward~F.~Schlafly}\orcidlink{0000-0002-3569-7421},}
\author[2]{{David~Schlegel},}
\author[39,40]{{Michael~Schubnell},}
\author[41]{{Hee-Jong~Seo}\orcidlink{0000-0002-6588-3508},}
\author[2]{{Joseph~Silber}\orcidlink{0000-0002-3461-0320},}
\author[8]{{David~Sprayberry},}
\author[1]{{Julien~Taran},}
\author[40]{{Gregory~Tarl\'{e}}\orcidlink{0000-0003-1704-0781},}
\author[8]{{Benjamin~A.~Weaver},}
\author[42,12]{{Martin~White}\orcidlink{0000-0001-9912-5070},}
\author[43]{{Michael~J.~Wilson},}
\author[28]{{Zhimin~Zhou}\orcidlink{0000-0002-4135-0977},}
\author[28]{{Hu~Zou}\orcidlink{0000-0002-6684-3997}.}

\affiliation[1]{IRFU, CEA, Universit\'{e} Paris-Saclay, F-91191 Gif-sur-Yvette, France}
\affiliation[2]{Lawrence Berkeley National Laboratory, 1 Cyclotron Road, Berkeley, CA 94720, USA}
\affiliation[3]{Physics Dept., Boston University, 590 Commonwealth Avenue, Boston, MA 02215, USA}
\affiliation[4]{Aix Marseille Université, CNRS, CNES, LAM, 38 rue Frédéric Joliot-Curie, 13388 Marseille cedex 13, France}
\affiliation[5]{Department of Physics \& Astronomy, University College London, Gower Street, London, WC1E 6BT, UK}
\affiliation[6]{Department of Physics and Astronomy, The University of Utah, 115 South 1400 East, Salt Lake City, UT 84112, USA}
\affiliation[7]{Instituto de F\'{\i}sica, Universidad Nacional Aut\'{o}noma de M\'{e}xico,  Cd. de M\'{e}xico  C.P. 04510,  M\'{e}xico}
\affiliation[8]{NSF NOIRLab, 950 N. Cherry Ave., Tucson, AZ 85719, USA}
\affiliation[9]{Department of Physics \& Astronomy and Pittsburgh Particle Physics, Astrophysics, and Cosmology Center (PITT PACC), University of Pittsburgh, 3941 O'Hara Street, Pittsburgh, PA 15260, USA}
\affiliation[10]{Kavli Institute for Particle Astrophysics and Cosmology, Stanford University, Menlo Park, CA 94305, USA}
\affiliation[11]{SLAC National Accelerator Laboratory, Menlo Park, CA 94305, USA}
\affiliation[12]{University of California, Berkeley, 110 Sproul Hall \#5800 Berkeley, CA 94720, USA}
\affiliation[13]{Departamento de F\'isica, Universidad de los Andes, Cra. 1 No. 18A-10, Edificio Ip, CP 111711, Bogot\'a, Colombia}
\affiliation[14]{Observatorio Astron\'omico, Universidad de los Andes, Cra. 1 No. 18A-10, Edificio H, CP 111711 Bogot\'a, Colombia}
\affiliation[15]{Fermi National Accelerator Laboratory, PO Box 500, Batavia, IL 60510, USA}
\affiliation[16]{NRC Herzberg Astronomy and Astrophysics, 5071 West Saanich Road, Victoria, BC V9E 2E7, Canada}
\affiliation[17]{Center for Cosmology and AstroParticle Physics, The Ohio State University, 191 West Woodruff Avenue, Columbus, OH 43210, USA}
\affiliation[18]{Department of Physics, The Ohio State University, 191 West Woodruff Avenue, Columbus, OH 43210, USA}
\affiliation[19]{The Ohio State University, Columbus, 43210 OH, USA}
\affiliation[20]{Department of Physics, Southern Methodist University, 3215 Daniel Avenue, Dallas, TX 75275, USA}
\affiliation[21]{Sorbonne Universit\'{e}, CNRS/IN2P3, Laboratoire de Physique Nucl\'{e}aire et de Hautes Energies (LPNHE), FR-75005 Paris, France}
\affiliation[22]{Departament de F\'{i}sica, Serra H\'{u}nter, Universitat Aut\`{o}noma de Barcelona, 08193 Bellaterra (Barcelona), Spain}
\affiliation[23]{Institut de F\'{i}sica d’Altes Energies (IFAE), The Barcelona Institute of Science and Technology, Campus UAB, 08193 Bellaterra Barcelona, Spain}
\affiliation[24]{Department of Astronomy, The Ohio State University, 4055 McPherson Laboratory, 140 W 18th Avenue, Columbus, OH 43210, USA}
\affiliation[25]{Instituci\'{o} Catalana de Recerca i Estudis Avan\c{c}ats, Passeig de Llu\'{\i}s Companys, 23, 08010 Barcelona, Spain}
\affiliation[26]{Department of Physics and Astronomy, Siena College, 515 Loudon Road, Loudonville, NY 12211, USA}
\affiliation[27]{Department of Physics and Astronomy, University of Sussex, Brighton BN1 9QH, U.K}
\affiliation[28]{National Astronomical Observatories, Chinese Academy of Sciences, A20 Datun Rd., Chaoyang District, Beijing, 100012, P.R. China}
\affiliation[29]{Departamento de F\'{i}sica, Universidad de Guanajuato - DCI, C.P. 37150, Leon, Guanajuato, M\'{e}xico}
\affiliation[30]{Instituto Avanzado de Cosmolog\'{\i}a A.~C., San Marcos 11 - Atenas 202. Magdalena Contreras, 10720. Ciudad de M\'{e}xico, M\'{e}xico}
\affiliation[31]{California Institute of Technology, Cahill Center for Astrophysics, Pasadena, CA, USA}
\affiliation[32]{Aix Marseille Université, CNRS/IN2P3, CPPM, 163 avenue de Luminy, 13288 Marseille cedex 09, France}
\affiliation[33]{Universit\'{e} Clermont-Auvergne, CNRS, LPCA, 63000 Clermont-Ferrand, France}
\affiliation[34]{Department of Physics, Kansas State University, 116 Cardwell Hall, Manhattan, KS 66506, USA}
\affiliation[35]{Department of Physics and Astronomy, Sejong University, Seoul, 143-747, Korea}
\affiliation[36]{CIEMAT, Avenida Complutense 40, E-28040 Madrid, Spain}
\affiliation[37]{Department of Astronomy \& Physics and Institute for Computational Astrophysics, Saint Mary’s University, 923 Robie Street, Halifax, NS B3H 3C3, Canada}
\affiliation[38]{Space Telescope Science Institute, 3700 San Martin Drive, Baltimore, MD 21218, USA}
\affiliation[39]{Department of Physics, University of Michigan, Ann Arbor, MI 48109, USA}
\affiliation[40]{University of Michigan, Ann Arbor, MI 48109, USA}
\affiliation[41]{Department of Physics \& Astronomy, Ohio University, Athens, OH 45701, USA}
\affiliation[42]{Department of Physics, University of California, Berkeley, 366 LeConte Hall MC 7300, Berkeley, CA 94720-7300, USA}
\affiliation[43]{Institute for Computational Cosmology, Department of Physics, Durham University, South Road, Durham DH1 3LE, UK}

\emailAdd{vanina.ruhlmann-kleider@cea.fr, christophe.yeche@cea.fr}

\abstract{Lyman break galaxies (LBGs) are promising probes for clustering measurements at high redshift, $z>2$, a region only covered so far by Lyman-$\alpha$ forest measurements. In this paper, we investigate the feasibility of selecting LBGs 
by exploiting the existence of a strong deficit of flux shortward of the Lyman limit, due to various absorption processes along the line of sight. The target selection relies on deep imaging data from the HSC and CLAUDS surveys in the $g,r,z$ and $u$ bands, respectively, with median depths reaching 27 AB in all bands.
The selections were validated by several dedicated spectroscopic observation campaigns with DESI. Visual inspection of spectra has enabled us to develop an automated spectroscopic typing and redshift estimation algorithm specific to LBGs. Based on these data and tools, we assess the efficiency and purity of target selections optimised for different purposes.
Selections providing a wide redshift coverage
retain $57\%$ of the observed targets after spectroscopic confirmation with DESI, and provide an efficiency for LBGs of $83\pm3\%$, for a purity of the selected LBG sample of $90\pm2\%$. This would deliver a confirmed LBG density of $\sim 620$ deg$^{-2}$ in the range $2.3<z<3.5$ for a $r$-band limiting magnitude $r<24.2$.
Selections optimised for high redshift efficiency  
retain $73\%$ of the observed targets after spectroscopic confirmation, with $89\pm4\%$ efficiency for $97\pm2\%$ purity. This would provide a confirmed LBG density of $\sim 470$ deg$^{-2}$ in the range $2.8<z<3.5$ for a $r$-band limiting magnitude $r<24.5$.
A preliminary study of the LBG sample 3d-clustering properties is also presented and used to estimate the LBG linear bias. A value of $b_{LBG} = 3.3 \pm 0.2 (stat.)$ is obtained for a mean redshift of 2.9 and a limiting magnitude in $r$ of 24.2, in agreement with results reported in the literature.
}


\begin{document}

\maketitle

\flushbottom
\clearpage


\section{Introduction}

Lyman break galaxies (LBGs) are young and actively star-forming galaxies that make up most of the population of star-forming galaxies at $z>1.5$.
Their spectra are characterized by a distinctive break 
at wavelengths shorter than the Lyman limit (912~\r{A}, rest-frame) and a decrement of flux shortward of the Lyman-$\alpha$ (Ly$\alpha$) spectral feature (1216~\r{A}, rest-frame), both due to absorption by neutral hydrogen. The Ly$\alpha$ spectral feature  
can be in absorption or emission, but in that case the emission is not as strong as in Lyman $\alpha$ emitters, which are lower mass, younger star-forming galaxies~\cite{2023PASA...40...52F}. LBG spectra also contain absorption lines at longer wavelengths, on a generally shallow continuum. These galaxies have been long studied to probe the properties of high redshift star-forming galaxies (e.g.~\citep{Steidel_1996,Steidel1999,Reddy2008,ReddySteidel2009,Hildebrandt2009})
and, more recently, to improve our understanding of galaxy populations during the epoch of reionisation (e.g.~\citep{Steidel2018}). They have been mostly studied at redshifts above 2 where the Lyman limit lies in the optical wavelength range.

Due to the lack of strong spectroscopic features at observed-frame optical wavelengths, except for a possible Ly$\alpha$ emission, LBGs require long exposure times to confidently assess their type and measure their redshifts, but high target densities can be reached. Consequently, they appear to be promising tracers to carry out 3d clustering measurements in the redshift range $2<z<4$, which until now has only been probed by Ly$\alpha$ forest measurements from high redshift quasars (QSOs)~\citep{eBOSSCosmo2021}. This remains the case for the ongoing survey by the Dark Energy Spectroscopic Instrument (DESI). However, for the DESI-II project - the upgrade of DESI currently under preparation~\citep{Schlegel2022} - LBGs are one of the main tracers, with expected densities exceeding that of the DESI QSO sample at $z>2.1$ by at least one order of magnitude.

DESI is a robotic, fiber-fed, highly multiplexed spectroscopic instrument that operates on the Mayall 4-meter telescope at Kitt Peak National Observatory~\citep{DESIOverview2022}. 
DESI can obtain simultaneous spectra of almost 5000 objects over a $\sim 3$ degree field~\citep{2016arXiv161100037D,Silber2023,DESIcorrector2023}. It is currently conducting a five-year survey of about a third of the sky, to obtain spectra from about 40 million galaxies and quasars~\citep{2016arXiv161100036D}. For the latter, DESI plans to observe 2 million QSOs at $0.9 < z < 2$ as direct tracers of the matter distribution and 0.8 million at $z > 2.1$ 
to probe the underlying matter distribution traced by the Ly$\alpha$ forest.
The preliminary DESI target selection for QSOs~\citep{QSO_TS2020} was tested extensively during the survey validation phase of DESI~\citep{EDR, 2024AJ....167...62A}  to define an optimised target selection for the main survey, as described in~\cite{QSO_TS_SV2023}. The final selection, based on an initial target density of 310~deg$^{-2}$, 
allows DESI to select more than 200 QSOs per sq. deg. (including 60 quasars with $z > 2.1$). Measurements of the baryon acoustic oscillations from the first year DESI sample of galaxies and quasars can be found in~\cite{DESI2024.III.KP4, DESI2024.IV.KP6} and the corresponding cosmological results are described in~\cite{DESI2024.VI.KP7A}. 

DESI-II aims to observe galaxies at higher densities and at higher redshifts than DESI, to serve as a pathfinder for a larger Stage-5 spectroscopic survey that would measure hundreds of millions of redshifts~\cite{Schlegel2022}. Of the different programs of DESI-II, LBGs at $2<z<4.5$ represent one third of the total expected target sample. 
DESI-II will operate with the same Mayall Telescope at Kitt Peak, either with the existing instrument or an upgraded one. As LBGs require most of the observing time of the instrument, they would highly benefit from an upgraded instrument~\cite{Schlegel2022}.

The aim of this paper is to describe a possible selection of LBG targets for DESI-II, based on a combination of imaging from the Hyper Suprime-Cam (HSC) Subaru Strategic Program (HSC-SSP) and the CFHT Large Area U-band Deep Survey (CLAUDS). The efficiency and purity of that selection were measured in a series of pilot surveys conducted with DESI. The outline of the paper is as follows. The target selection followed for the first campaign of observations is described in section~\ref{sec:firstTS}. Its validation based on visual inspection of spectra is presented in section~\ref{sec:validTS}. 
Methods for automated LBG typing and redshift measurements are discussed in section~\ref{sec:tools}. Extended target selections tested in further pilot surveys are presented in section~\ref{sec:new_selections} and their performance, established using the automated algorithm, is detailed in section~\ref{sec:perfTS}. Section~\ref{sec:clustering} presents clustering measurements for the LBG sample collected during the pilot surveys. We conclude in section~\ref{sec:conclusion}.
Throughout the paper, all magnitudes are corrected for Galactic extinction, using dust maps from~\cite{1998ApJ...500..525S}.

\section{Target selection of LBGs in first observation campaign}
\label{sec:firstTS}

The most noticeable feature of LBG spectra is the flux deficit (or 'break`) shortward of the 912~\r{A} Lyman limit in the emitter rest-frame, due to absorption along the line of sight by neutral hydrogen rich stellar atmospheres and interstellar continuum photoelectric absorption. In broadband optical imaging, this feature can be exploited when the Lyman limit is observed in the optical domain, i.e. for redshifts around 2.3 and beyond, depending on the filters used. 

In addition, resonant line scattering from the Lyman series along the line-of-sight suppresses flux shortward of the 1216~\r{A} Ly$\alpha$ line in the rest-frame,
for galaxies at redshift greater than 2 and
the Ly$\alpha$ forest contribution dominates resonant line scattering for galaxies in the range $2 < z < 4$~\cite{1995ApJ...441...18M}. In broadband optical imaging, this feature can be exploited as soon as the Ly$\alpha$ forest is observed in the optical range, i.e. for redshifts greater than about 1.5, depending on the filters used. 

The Lyman continuum dropout due to the cumulative effect of absorbing material along the line of sight has been long advocated as a way to identify high-z galaxies using sufficiently deep broadband imaging~\citep{Meier1976, 1995ApJ...441...18M} and an adequate combination of filters depending on the redshift range to be targeted~\cite{Steidel_1996, Steidel1999}. This is the path we follow in this paper.

\subsection{CLAUDS and HSC imaging data}
The LBG target selection relies on the deep $gri$ HSC imaging~\cite{2019PASJ...71..114A} completed by the $U$-band imaging from the CLAUDS survey~\citep{CLAUDS19}, which was designed to follow up the HSC deep fields to similar depth with MegaCam at CFHT. 
The four deep fields of HSC are E-COSMOS, XMM-LSS, ELAIS-N1 and DEEP2-3. Two of them include ultra-deep fields, COSMOS (in E-COSMOS) and SXDS (in XMM-LSS). All are in well studied areas of the sky, thus providing ideal fields to test-bench target selections for new tracers foreseen for future cosmology projects. This work uses the E-COSMOS (referred to as COSMOS throughout this paper) and XMM-LSS fields.

The HSC imaging of the deep and ultra-deep fields
has reached unprecedented depth, e.g. 
$i_{lim}\sim 27.1$~AB and $27.7$~AB, respectively 
($5\sigma$ for point sources)~\citep{2019PASJ...71..114A}.
In these fields, CLAUDS provide $U$-band data of comparable depth and areal coverage, with a median depth of $U=27.1$~AB and $27.7$~AB in the deep and ultra-deep fields, respectively ($5\sigma$ in $2"$ apertures)~\citep{CLAUDS19}. These correspond to a median depth of $\sim$ 27.5 AB measured as 5$\sigma$ point-source limits.

The CLAUDS programme used two filters in the $U$-band, the original MegaCam $u^*$-filter and an upgraded $u$-filter installed in 2014, which has better throughput and is physically larger, allowing the entire MegaCam mosaic of 40 CCDs (instead of 36) to be illuminated. The XMM-LSS field was imaged with the $u^*$ filter, while both $u$ and $u^*$ filters were used for the COSMOS field~\citep{CLAUDS19}. For the latter, 
the LBG target selections described in this paper are defined either from  the $u$-filter or from a combination of the $u$ and $u^*$ filters, depending on the observation campaign {(see table~\ref{tab:cuts}).

The pilot surveys conducted with DESI to test the LBG target selections included three observation campaigns, two on the COSMOS field in 2021 and 2023, and one on the XMM-LSS field in 2022 (see table~\ref{tab:pilot}). The target selections for the first observation campaign are described in the next subsections. Those applied during the second and third campaigns are presented in section~\ref{sec:new_selections}.

\subsection{\texorpdfstring{$u$}{u}-dropout target selection}
\label{sec:u_dropout}

\begin{figure} [t]
\centering
\includegraphics[width=\textwidth]{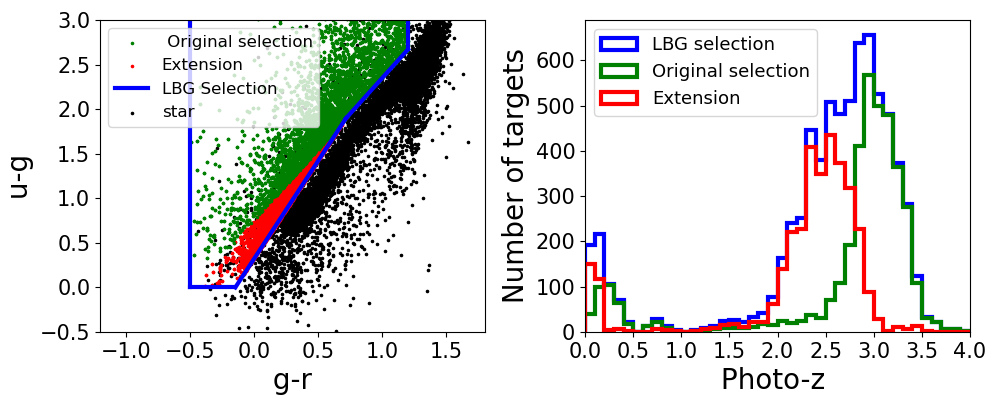}
\caption{{\it Left:} Color selection  of LBGs (blue polygon) applied in the first DESI pilot survey carried out on the COSMOS field in 2021 (see first two lines in table~\ref{tab:cuts}). Green dots stand for the original $U$-dropout selection advocated by~\cite{CLAUDS19}, red dots correspond to an extension to enhance the target density in the range $2 < z <2.8$. Black dots are stars identified by their point-source morphology in HSC catalogs.  {\it Right:} Photometric redshift distributions of the three target selections applied to the CLAUDS catalogs, based on the \textsc{LePHARE} photometric redshift code.
}
\label{fig:tmg}
\end{figure} 
The first target selection was designed for the COSMOS field, using CLAUDS data from the $u$-filter.
The selection keeps only objects in the CLAUDS catalog with a magnitude between 22.5 and 24.5 in the $r$-band and then applies a $u$-dropout selection. This technique consists of using the flux decrement blueward of the Lyman limit or Ly$\alpha$ spectral feature, due to absorption by neutral hydrogen.  In practice, a $u$-dropout means that objects are selected for their lack of detection in the $u$-band, while having significant flux in one or more redder bands.
Note that for galaxies in the range $2<z<4$, the $u$-dropout originates only from the flux decrement blueward of the Ly$\alpha$ spectral line up to $z\sim2.3$, above which the Lyman break becomes visible in the $u$-band and also contributes.

Inspired by the original selections described in~\cite{CLAUDS19}, we use a color-color box extended so as to increase the redshift coverage of the LBG sample. Figure~\ref{fig:tmg} shows the original and extended selection boxes in the $u-g$ vs $g-r$ diagram, which appear well separated from the star locus, with stars identified by their point-source morphology in the HSC catalogs~\cite{HSC2018}. The corresponding distributions of the target photometric redshifts provided by the CLAUDS collaboration using the \textsc{LePHARE} code~\citep{Arnouts99,2002MNRAS.329..355A,2006A&A...457..841I} are shown in the right-hand plot. 
This figure shows how effective the extension is in the $2.0<z<2.8$ range.
The above selection is divided into two sub-selections, called TMG and BXU selections whose cuts are detailed in the first two lines of table~\ref{tab:cuts}. In figure~\ref{fig:tmg}, and all subsequent figures showing the color box of the $u$-dropout selection of the first campaign, 
an OR of the TMG and BXU color-color boxes is plotted.

\begin{table}[t]
\centering
{\scriptsize
\begin{tabular}{|c|c|l|}\hline
\multicolumn{3}{|l|}{COSMOS: TMG $u$-dropout ($22.5<r<23.75$)} \\ \hline
$u-g>0.3$ & $-0.5<g-r<1.0$ & $(u-g>2.2(g-r)+0.32) \cup (u-g>0.9 \cap u-g>1.6(g-r)+0.75)$ \\
\hline
\multicolumn{3}{|l|}{COSMOS: BXU $u$-dropout ($23.75<r<24.5$)} \\ \hline     
$u-g>0$ & $-0.5<g-r<1.2$ & $(u-g>2.2(g-r)+0.32) \cup (u-g>0.9 \cap u-g>1.6(g-r)+0.75)$ \\ \hline
\multicolumn{3}{|l|}{COSMOS: $g$-dropout ($22.5<i<25.5$)} 
\\ \hline     
$g-r>1$ & $-1.5<r-i<1.0$ & $(g-r>1.5(r-i)+0.8)$ \\
\hline
\multicolumn{3}{|l|}{XMM-LSS: $u^*$-dropout ($22.7<r<24.2$)} \\ \hline 
$u^*-g>0.3$ & $0<g-r<0.8$ & $(u^*-g>2.0(g-r)+0.42) \cup (u^*-g>1.6(g-r)+0.55)$ \\ \hline
\multicolumn{3}{|l|}{XMM-LSS: low-$z$ extension ($22.7<r<24.2$)} \\ \hline 
$u^*-g>0.45$ & $0<g-r<0.8$ & $(u^*-g>3.5(g-r)+0.12)\cap$ (outside the $u^*$-dropout color-box) \\  \hline
\multicolumn{3}{|l|}{XMM-LSS: faint $u^*$-dropout ($24.2<r<24.5$)} \\ \hline 
$u^*-g>0.3$ & $0<g-r<0.8$ & $(u^*-g>2.0(g-r)+0.42) \cup (u^*-g>1.6(g-r)+0.55)$ \\ \hline
\multicolumn{3}{|l|}{COSMOS: $U$-dropout ($22.7<r<24.5$)} \\ \hline 
$U-g>0.3$ & $0<g-r<0.8$ & $(U-g>2.2(g-r)+0.32) \cup (U-g>1.6(g-r)+0.75)$ \\ \hline
\end{tabular}
}
 \caption{LBG target selection cuts studied in this paper. In the last line, $U$ denotes an appropriate combination of the $u$ and $u^*$ filters used by CLAUDS.
 }
\label{tab:cuts}
\end{table}

\begin{table}[t]
\centering
{\footnotesize
\begin{tabular}{|c|l|c|c|c|c|c|c|c|}\hline
field & sample & year & selection & $T_{eff}$ & $N_{targets}$ & $N_{spectra}$ & $N_{VI}$ & sections\\ \hline
COSMOS & 1 (tile 80871)  & 2021 & TMG & 5 h & 1982 & 811 & 811 & \ref{sec:u_dropout},~\ref{sec:validTS} \\
COSMOS & 2 (tile 80872)  & 2021 & TMG & 5 h & 1047 & 508 & - & \ref{sec:u_dropout},~\ref{sec:validTS} \\ 
COSMOS & 3 (tile 80871)  & 2021 & BXU & 5 h & 12103 & 1045 & 617 & \ref{sec:u_dropout},~\ref{sec:validTS} \\
COSMOS & 4 (tile 80872)  & 2021 & BXU & 5 h & 10859 & 1426 & - & \ref{sec:u_dropout},~\ref{sec:validTS} \\  
COSMOS & 5 (tile 80871) & 2021 & $g$-dropout & 5 h & 26470 & 698 & - & \ref{sec:g_dropout},~\ref{sec:g_eff} \\
COSMOS & 6 (tile 80872) & 2021 & $g$-dropout & 5 h & 25654 & 914 & - & \ref{sec:g_dropout},~\ref{sec:g_eff}  \\
XMM-LSS & 7 (rosette) & 2022 & $u^*$-dropout & 4 h & 3047 & 858 & 345 & \ref{sec:refinedTS},~\ref{sec:perfTS} \\ 
XMM-LSS & 8 (rosette) & 2022 & $u^*$-dropout & 2 h & 3047 & 1523 & 50 &\ref{sec:refinedTS},~\ref{sec:perfTS} \\ 
XMM-LSS & 9 (rosette) & 2022 & low-$z$ ext. & 4 h & 3000 & 702 & 50 & \ref{sec:refinedTS},~\ref{sec:perfTS} \\ 
XMM-LSS & 10 (rosette) & 2022 & low-$z$ ext. & 2 h & 3000 & 1427 & - & \ref{sec:refinedTS},~\ref{sec:perfTS} \\ 
XMM-LSS & 11 (rosette) & 2022 & faint $u^*$-dropout & 3 h & 5617 & 2741 & 50 & \ref{sec:refinedTS},~\ref{sec:perfTS} \\ 
COSMOS & 12 (rosette) & 2023 & $U$-dropout & 4 h & 6249 & 2183 & - & \ref{sec:refinedTS},~\ref{sec:clamato} \\     
\hline     
\end{tabular}
}
\caption{Pilot surveys conducted on various fields with DESI to assess the performance of different LBG target selections, based on the dropout technique described in this paper. The selections are defined in table~\ref{tab:cuts}.
The column labelled $N_{spectra}$ shows the number of spectra satisfying \texttt{COADD\_FIBERSTATUS==0}, which means that there were no hardware or observing issues for
all input data for those spectra~\cite{EDR}.
The column labelled $N_{VI}$ shows the number of spectra which were visually inspected.
The difference between the numbers of targets and spectra mainly reflects the observation time allotted to each pilot survey. Overall, a sample of 14836 spectra was collected and 1923 of these were visually inspected. The last column cites the sections which describe the selections and their performance.}
\label{tab:pilot}
\end{table}

\subsection{\texorpdfstring{$g$}{g}-dropout target selection}
\label{sec:g_dropout}

The $u$-dropout selection, based on the three $ugr$ bands presented in the previous paragraph, allows LBGs to be selected in a redshift range between 2.5 and 3.5. In a similar way, another selection was developed~\cite{Wilson19} based on the three $gri$ bands to select higher redshifts, as can be seen in the left-hand plot of figure~\ref{fig:color_box_g_dropout}.

The goal is to use again the flux decrement blueward of the Lyman limit or  Ly$\alpha$ spectral feature due to absorption by neutral hydrogen, but selecting objects 
for their lack of a detection in the $g$-band. 
Given the definition of the HSC $g$-band, the $g$-dropout originates only from the flux decrement blueward of the Ly$\alpha$ spectral line up to $z\sim3.3$, above which the Lyman break becomes visible in the $g$-band and also contributes.

The $g$-dropout selection cuts are detailed in table~\ref{tab:cuts}. As a result, the selected LBGs  have a redshift in the range between 3.5 and 4.5, as shown by the right-hand plot in figure~\ref{fig:color_box_g_dropout}.
Due to their larger redshift, these LBGs are much fainter. Objects in this selection are thus required to have $r<25.5$ to achieve a sufficiently high density, between 500 and 1000~deg$^{-2}$.

\begin{figure} [t]
\centering
\includegraphics[width=1.0\textwidth]{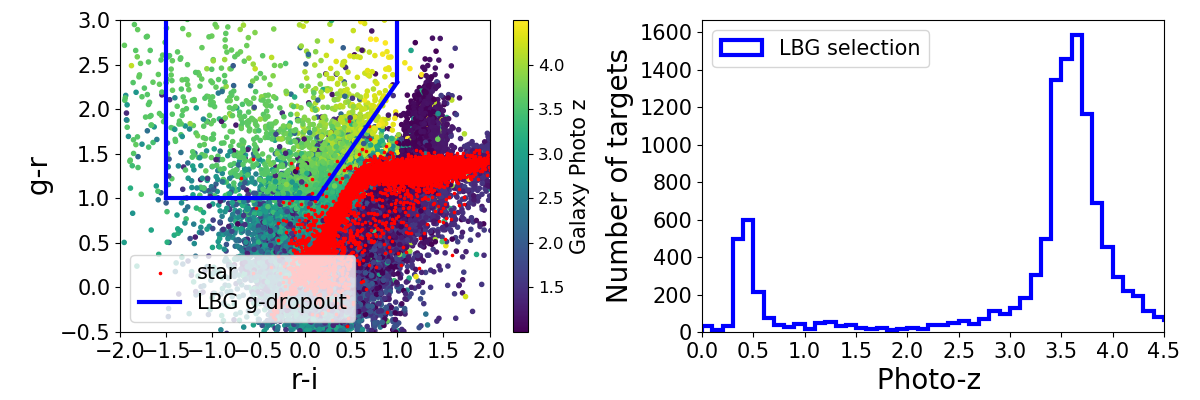} 
\caption{ {\it Left:} The blue line is the LBG  $g$-dropout selection box applied to the COSMOS field in the 2021 DESI pilot survey (see table~\ref{tab:cuts}). Red dots are HSC stars and medium size dots colored by photometric redshift based on the \textsc{LePHARE} code are HSC galaxies. {\it Right:}  Photometric redshift distribution for the $g$-dropout selection. 
}
\label{fig:color_box_g_dropout}
\end{figure}

\subsection{DESI observations}
The above target selections were tested in a pilot survey carried out by DESI in 2021 on the COSMOS field with an effective exposure time of $\sim$ 5 hours. 
The numbers of spectra collected during this pilot survey are given in the first six lines in table~\ref{tab:pilot}, separately for each of the three selections and for the two samples that were targeted for each selection. 

Data from the above LBG pilot survey are part of the Early Data Release (EDR) recently made public by the DESI collaboration~\citep{EDR} (see appendix B35).

\section{Validation of the \texorpdfstring{$u$}{u}-dropout selection in first campaign}
\label{sec:validTS}

\begin{figure} [t]
\centering
\includegraphics[width=\textwidth]{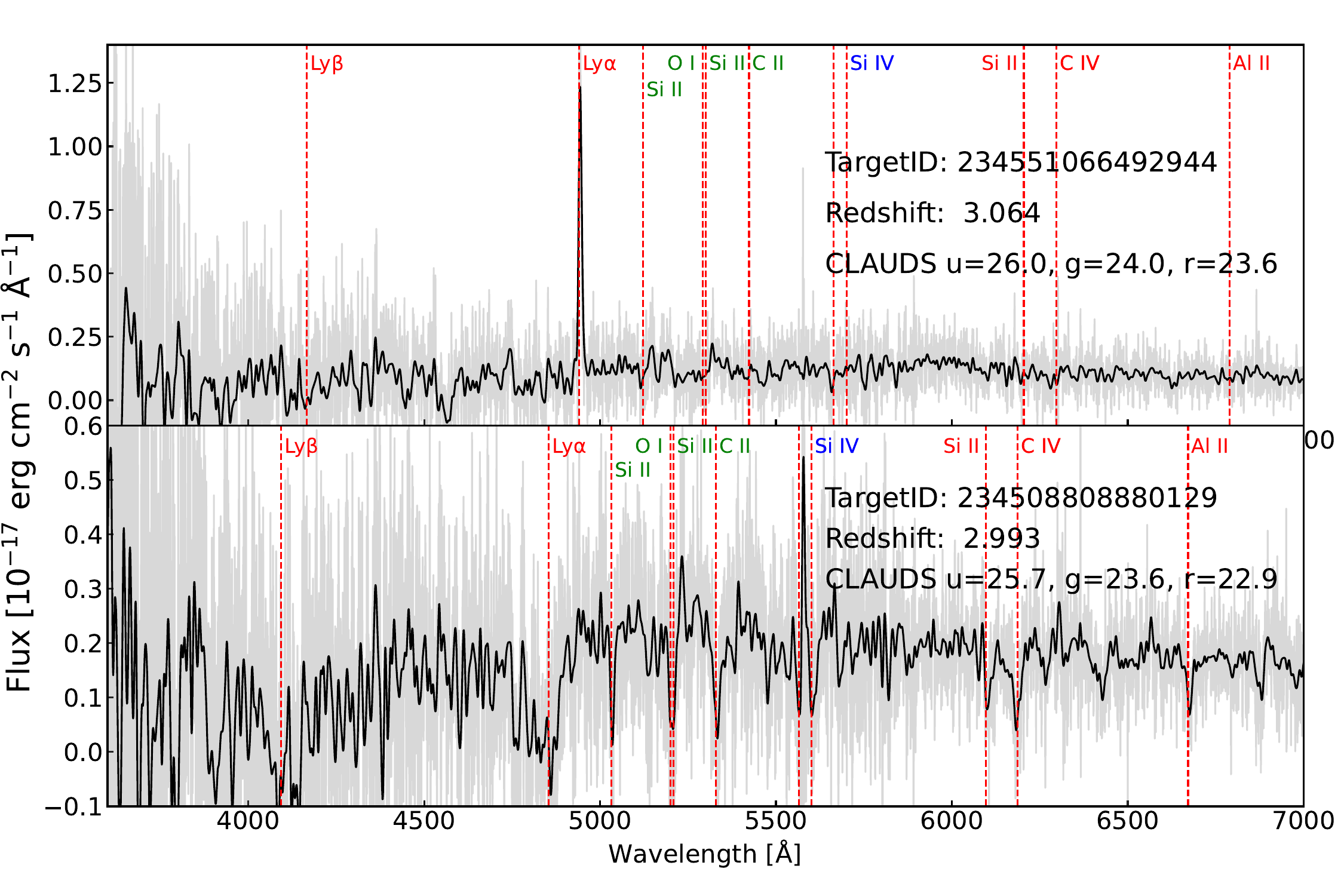}
\caption{Two spectra of LBG targets from the $u$-dropout TMG selection measured during the pilot survey of 2021 on the COSMOS field. The raw spectra are in grey, the smoothed ones are in black. A Gaussian smearing with a standard deviation of 4 pixels is used. {\it Top:} a significant Ly$\alpha$ emission is present, hence absorption lines are weak. {\it Bottom:} a broad Ly$\alpha$ absorption is present as well as a sequence of quite visible absorption lines redward from the Ly$\alpha$ line. Lines labelled in green (resp. blue) are low (resp. high) ionization interstellar metal lines which are used as main redshift indicators during visual inspection.
}
\label{fig:spectra}
\end{figure}

Spectra of the fiber-assigned LBG targets from the $u$-dropout selections were measured by the DESI 3-arm spectrographs which record the source light from $3600-9800$~\r{A} with a spectral resolution that ranges from 2000 to 5000~\citep{DESIOverview2022}. Due to the redshift range of LBG targets, most spectra have their distinctive spectral features in the blue channel of the spectrographs ($3600-5930$~\r{A}) whose resolution ranges between 2000 and 3000. During the first pilot survey, several exposures were taken for a total effective time of $\sim$ 5 hours. The raw spectra were then treated (i.e. pre-processed, wavelength calibrated, extracted, flat-fielded, sky background subtracted and flux calibrated) by the standard DESI spectroscopic reduction pipeline described in~\cite{Guy2023}. In this work, we use co-added calibrated spectra from different exposures for the same objects.

\subsection{Visual inspection}
\label{sec:vi}
A visual inspection (VI) campaign was carried out on the complete sample of 811 spectra of the $u$-dropout TMG selection in tile 80871 (see table~\ref{tab:pilot}). This is the only sample that has undergone full visual inspection and is therefore suitable for quantitative assessment of the content of the selected sample of LBG targets. Spectra were scanned independently by three inspectors and inconsistencies on type or redshift were solved after rescan by one of the inspectors. Examples of spectra  are shown in figure~\ref{fig:spectra}. Spectra fall into two different classes, with or without a Ly$\alpha$ emission. As described in~\cite{Shapley2003}, we observe that in the first case, the stronger the Ly$\alpha$ emission, the weaker the absorption lines, while in the second case, absorption lines are generally visible redward from a marked Ly$\alpha$ absorption. 

These criteria were used during VI to confirm the LBG type and to 
assign redshifts. In particular, for LBGs with no emission, we rely on the strongest absorption features for both type and redshift, namely
\ion{Si}{II} $\lambda 1260$, \ion{O}{I} $\lambda 1302$+\ion{Si}{II} $\lambda 1304$, \ion{C}{II} $\lambda 1335$ (green labels in figure~\ref{fig:spectra}) and \ion{Si}{IV} $\lambda\lambda 1394,1403$ (blue label). Other absorption lines such as Ly$\beta$ $\lambda 1026$, \ion{S}{V}~$\lambda 1501$, \ion{Si}{II}~$\lambda 1527$, \ion{C}{IV}~$\lambda\lambda 1548,1551$, \ion{Fe}{II} $\lambda 1608$, \ion{Al}{II} $\lambda 1670$, if visible, bring an additional confirmation for the type. Note that \ion{Si}{II} $\lambda 1260$, \ion{O}{I} $\lambda 1302$+\ion{Si}{II} $\lambda 1304$ and \ion{C}{II} $\lambda 1335$ are low-ionization resonance interstellar metal lines while \ion{Si}{IV} $\lambda\lambda 1393,1403$ are high-ionization metal lines associated with ionized interstellar gas and P-Cygni stellar wind features. These lines have different physical origins, but their velocity offsets in the systemic rest-frame have a sufficiently small dispersion~\citep{Shapley2003} compared to the best redshift resolution we can achieve in VI, $\Delta z \sim 0.005$, so that we can take advantage of all of them to estimate the redshift by visual inspection.

On the other hand, when a Ly$\alpha$ emission is present, the velocity offset of the emission line compared to absorption ones is greater than the above desired redshift uncertainty (a mean $\Delta z \sim 0.008$ is found in \cite{Shapley2003}). In order to avoid a systematic bias in the VI redshift measurement, we use the same sequence of absorption lines as in the case of pure absorption LBGs to correct the estimate of the redshift that would be deduced from the Ly$\alpha$ emission maximum. This usually results in the redshift being fixed in the rising part of the emission peak.

\begin{figure} [t]
\centering
\includegraphics[width=\textwidth]{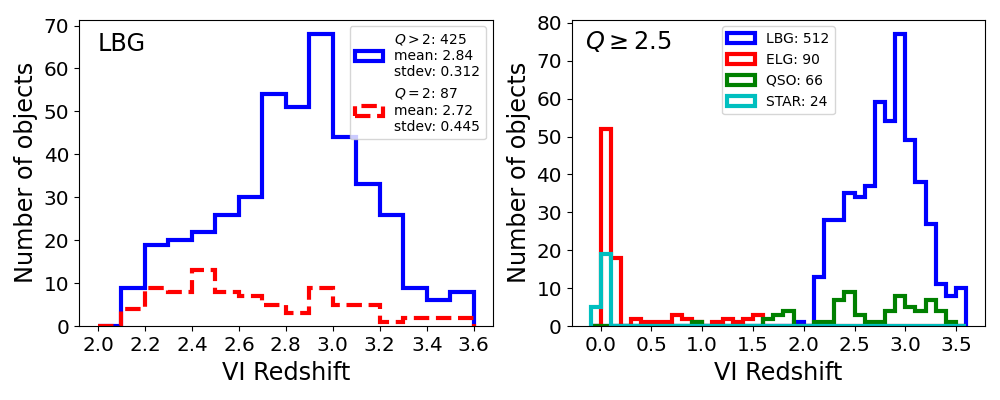}
\caption{VI results from the LBG pilot survey carried out by DESI in spring 2021 on the COSMOS field, 
for the whole sample of spectra from the TMG selection applied to tile 80871.
{\it Left:} VI redshift distribution of all LBG spectra with secure typing from VI. Spectra with VI quality of 2 have no secure redshift from VI. {\it Right:} VI redshift distribution of all objects with secure typing from VI. The LBG selection is contaminated by ELGs at low redshift and by QSOs at hight redshift.
}
\label{fig:VI_z}
\end{figure}

Following what was done for the DESI VI campaigns carried out during survey validation~\cite{2023ApJ...943...68L,2023AJ....165..124A}, each inspector assigned a VI quality flag to each spectrum, with the following meaning: $Q=3,4$ for secure redshifts, $Q=2$ for non robust ones and $Q<2$ for no redshift assignment i.e. for spectra compatible with noise or a weak continuum with no clear feature. The final VI quality flag is the mean of the three inspector flags in case of compatible answers or that of the final inspection in case a rescan was necessary to solve inconsistencies. 

The results of the VI campaign are as follows. Out of the 811 spectra, 696 have secure types, among which 598 have also secure redshifts, i.e. $Q\ge2.5$. 426 of these are LBGs 
and the 172 other spectra (29$\%$) are contaminants (88 ELGs, 65 QSOs and 19 STARS). The breakdown of the 426 LBG spectra by VI type is 
as follows: LBGs in absorption (111 spectra) or with a significant Ly$\alpha$ emisssion (101 spectra) represent $\sim$25$\%$ of the LBG sample each, while the remaining 50$\%$ have absorption lines redward from a moderate Ly$\alpha$ emission (214 spectra). This is in qualitative agreement with the findings of~\cite{Shapley2003}. 
The 98 spectra with secure type but no robust redshift $(Q=2)$ are mostly LBGs (86), among which 61 pure absorption ones. 
This illustrates the fact that, in the absence of Ly$\alpha$ emission, it is difficult to obtain a robust redshift, as Ly$\alpha$ absorption can be difficult to detect e.g. if the continuum is weak or if the spectrum is noisy. In these cases, different plausible Ly$\alpha$ absorptions may exist and it is not always possible to solve these degeneracies by VI with the help of the other absorption lines. An automated procedure appears to be a better way, as discussed in section~\ref{sec:tools}. 

The left-hand plot in figure~\ref{fig:VI_z} presents the VI redshift distribution of the LBG sample, separately for good quality spectra and spectra with secure type but no robust redshift. 
The bulk of the latter distribution is in the low redshift region where we experienced more difficulties to assign redshifts due to degeneracies between different possible solutions. Note also that the redshift distribution starts at $z\sim2.1$, below which it is not possible to obtain reliable redshifts with DESI for faint objects like LBGs, because the Ly$\alpha$ feature is 
in the low-throughput region of the blue spectrographs (see also section~\ref{sec:throughput}).

The right-hand plot of the figure compares the LBG VI redshift distribution with those of the contaminants. While QSOs are mainly at high redshift and overlap with the LBG sample, ELG contaminants appear to be mostly low redshift ones. This was used in our refined selections to reduce the ELG contamination, as discussed in section~\ref{sec:extension}. Visual inspection showed that part of the low redshift ELGs are Balmer break galaxies. 

The TMG sub-sample of spectra was the largest complete sample submitted to visual inspection in this work. However, when necessary, we resorted to VI on partial sub-samples of data (see table~\ref{tab:pilot}) e.g. to measure performance indicators, as reported in some places in this paper. 

\begin{figure} [t]
\centering
\includegraphics[width=\textwidth]{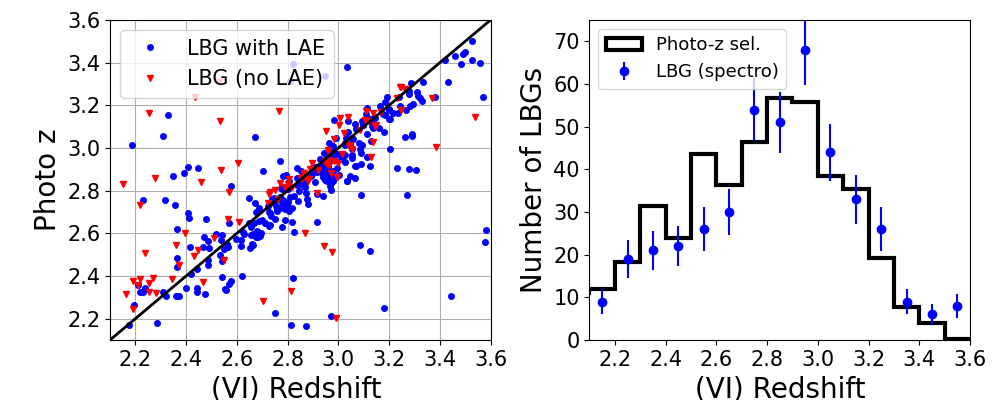} 
\caption{Results from the LBG pilot survey carried out by DESI in spring 2021 on the COSMOS field, 
for the whole sample of spectra from the TMG selection applied to tile 80871: comparison between VI redshifts and LePHARE photometric redshifts for the VI-typed LBGs with 
secure VI redshifts ($Q\ge 2.5$).
{\it Left:} Blue dots (resp. red triangles)  stand for LBGs with (resp. without) Ly$\alpha$ emission. {\it Right:} LBG spectroscopic redshift distribution from the visually inspected sample (blue dots) and normalized distribution of photometric redshifts for all LBG targets (black solid line).  }
\label{fig:redshifts}
\end{figure}

\subsection{Redshift range of the \texorpdfstring{$u$}{u}-dropout selection}

The VI redshifts of the good quality LBG sample discussed in the previous section range from 2.1 to 3.6, with a mean resdhift for the good quality spectra of 2.8 and a dispersion of 0.3 (see figure~\ref{fig:VI_z}).
In figure~\ref{fig:redshifts}, VI redshifts for that sample are compared to the~\textsc{LePHARE} photometric redshifts which were used to define the target selection.
The agreement between the spectroscopic and photometric redshifts is satisfactory despite the dispersion. 

As shown in the right-hand plot, there is a modest systematic shift between the mean values of the spectroscopic redshift distribution for the visually inspected sample and the photo-$z$ distribution for all targets. 
Indeed, for low-$z$ LBGs, the normalized photo-$z$ histogram (black) is above the blue dots corresponding to spectroscopic redshifts, and below the blue dots for redshift above 2.8. This may reflect the difficulty to visually assign LBG redshifts in DESI for $z<2.8$ and/or the quality of the photometric redshift. 

However, the difference is rather small and does not prevent {\sc LePHARE} photometric redshifts from being used to refine the target selection, which will be detailed in section~\ref{sec:new_selections}.
The new selections were those tested during the second and third campaigns on the XMM-LSS and COSMOS fields, reported in table~\ref{tab:pilot}.

\subsection{Results of the \texorpdfstring{$u$}{u}-dropout selection}
\label{sec:res_vi}

\begin{figure} [t]
\centering
\includegraphics[width=1.0\textwidth]{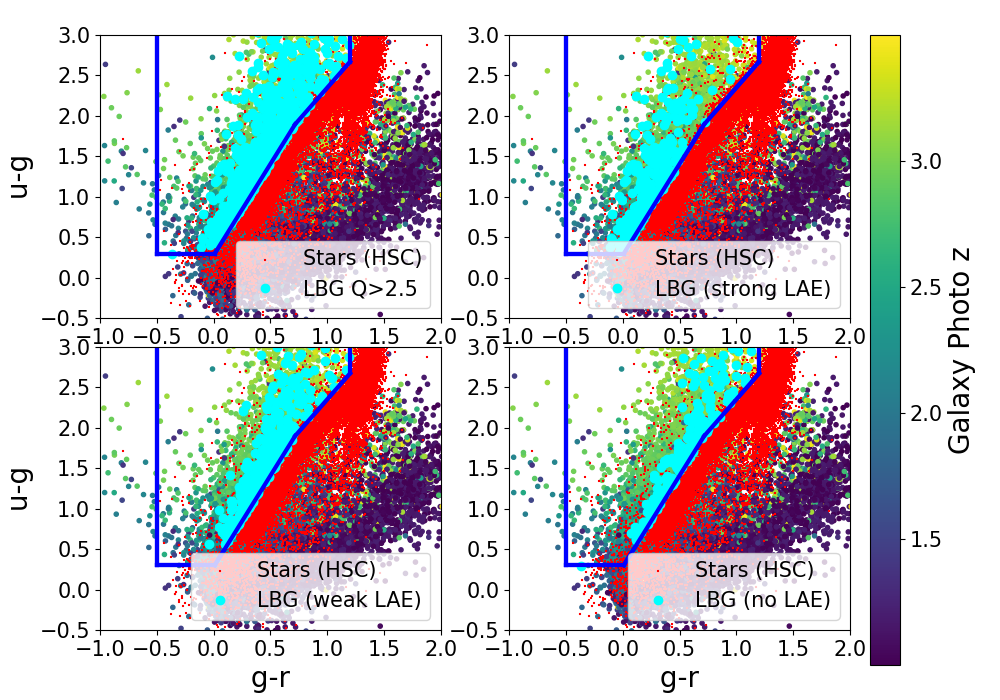} 
\caption{Results from the LBG pilot survey carried out by DESI in spring 2021 on the COSMOS field, tile 80871, for the TMG and BXU target selections. In all plots, the blue line is the LBG selection box, red dots are HSC stars and medium size dots colored by photometric redshift are HSC galaxies. The cyan circles represent LBGs identified by Visual Inspection. From left to right and from top to bottom, there are successively all LBGs with a good VI quality ($Q\ge 2.5$) and those among them with a strong Ly$\alpha$ emission line,  a weak Ly$\alpha$ emission line, and no Ly$\alpha$ emission line. 
}
\label{fig:lbg_colourbox}
\end{figure}
\begin{figure} [t]
\centering
\includegraphics[width=1.0\textwidth]{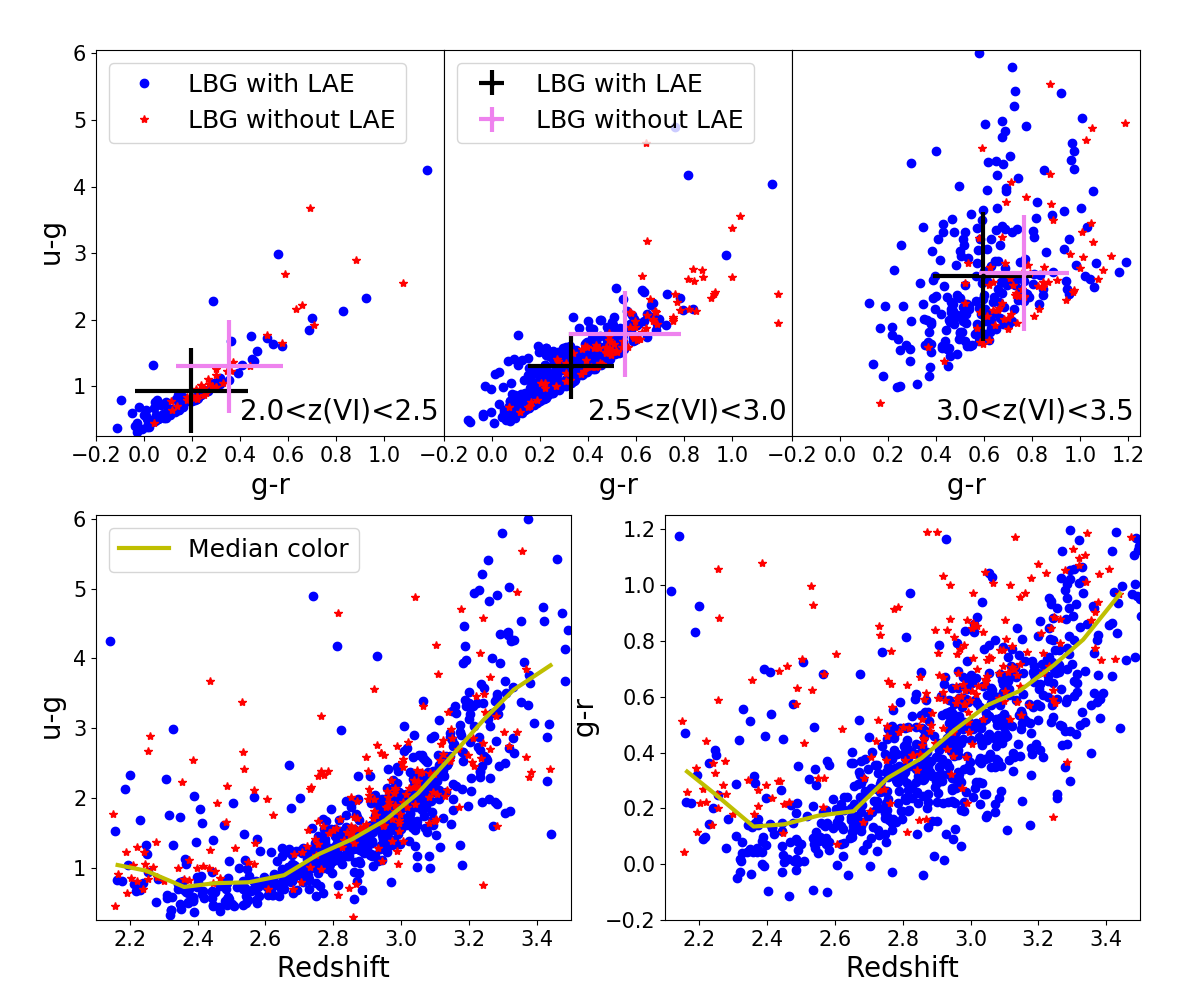} 
\caption{Results from the LBG pilot survey carried out by DESI in spring 2021 on the COSMOS field, tile 80871, for the TMG and BXU target selections. {\it Top:} distributions in the  selection box of VI-typed LBGs of good quality ($Q\ge 2.5$) with and without Ly$\alpha$ emission, shown in three bins of VI redshift. The black and pink crosses indicates the mean colors of the two populations and their statistical errors. {\it Bottom:} distributions of $u-g$ and $g-r$ colors as a function of VI redshift, shown separately for the same two sub-samples of LBGs. The green line is the median color of the total LBG sample.
}
\label{fig:lbg_lae}
\end{figure}

\begin{figure} [ht]
\centering
\includegraphics[width=\textwidth]{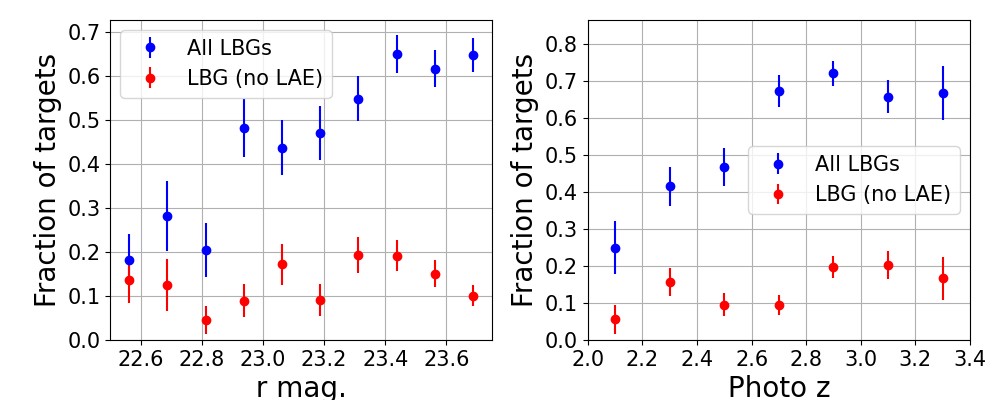} 
\caption{Results from the LBG pilot survey carried out by DESI in spring 2021 on the COSMOS field, for the whole sample of spectra from the TMG selection applied to tile 80871. Fractions of good quality VI-typed LBGs in the VIed target sample: blue (resp. red) dots stand for all LBGs (resp. LBGS with no Ly$\alpha$ emission). Evolution as a function of the HSC $r$-band magnitude ({\it left}) and the LePHARE photometric redshift ({\it right}).
}
\label{fig:fraction_lbg}
\end{figure}

\begin{figure} [t]
\centering
\includegraphics[width=1.0\textwidth]{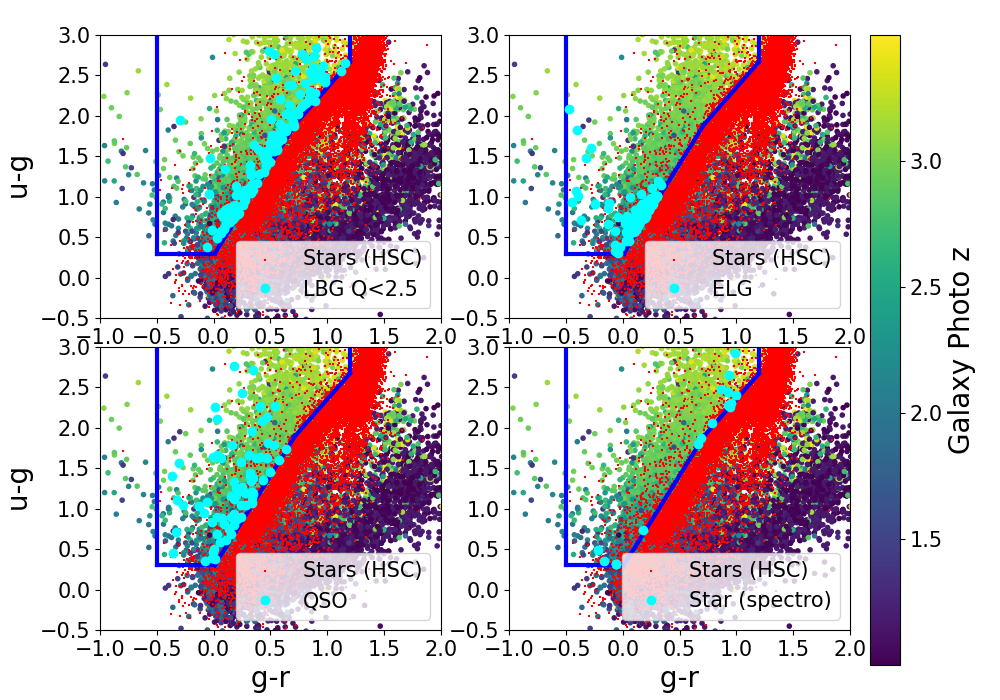} 
\caption{Results from the LBG pilot survey carried out by DESI in spring 2021 on the COSMOS field, tile 80871, for the TMG and BXU target selections. In all plots, the blue line is the LBG selection box, red dots are HSC stars and medium size dots colored by photometric redshift are HSC galaxies. The cyan circles represent contaminants identified by Visual Inspection. From left to right and from top to bottom, there are successively LBG targets with a low VI quality ($Q<2.5$), ELGs,  QSOs and stars. }
\label{fig:contaminant_colourbox}
\end{figure}

Figure~\ref{fig:lbg_colourbox} shows the location of the visually confirmed LBGs with a good quality ($Q\ge 2.5$) in the $(g-r,u-g)$ color plane. The stellar locus in this figure (and in figure~\ref{fig:contaminant_colourbox}) is derived from a selection of stars based on their point source morphology in HSC imaging catalogs~\cite{HSC2018}. More explicitly, we require the HSC \texttt{classification\_extendedness} flag for the $r$-band to be 0 to select point-like sources, and require in addition $r<23$ to ensure that morphological cuts provide true stars. 

Based on visual inspection, we divided these LBGs into three sub-samples: 
LBGs with a strong Ly$\alpha$ emission,  LBGs with a weak Ly$\alpha$ emission and LBGs with no Ly$\alpha$ emission. They represent respectively $24\%$, $50\%$, and $26\%$ of the confirmed LBG sample. 
The position in the color plot is significantly different for these three sub-samples. 
The weaker the Ly$\alpha$ emission, the closer LBGs are to the stellar locus.
In addition, LBGs with no or weak Ly$\alpha$ emission exhibit redder 
colors, on average, than LBGs with a strong Ly$\alpha$ emission, a trend observed whatever the redshift in the range $2.0<z<3.5$, as shown in figure~\ref{fig:lbg_lae}. This is in-line with results reported in the literature, e.g.~\cite{Shapley2003,2023PASA...40...52F}.
In order to favor LBGs with a pronounced Ly$\alpha$ emission line,  whose redshift is easier to determine,  we decreased the $g-r$ color cut to  $g-r<0.8$ in subsequent pilot surveys (see table~\ref{tab:cuts} and section~\ref{sec:refinedTS}).

Figure~\ref{fig:fraction_lbg} shows the variation of the number of good quality VI-typed LBGs relative to the total number of VIed TMG targets as a function of $r$ band magnitude and photometric redshift. 
The fraction of secure LBGs drops at brighter magnitudes and lower redshifts. As brighter objects are expected to be at lower redshifts, the two effects are likely to have the same origin. Part of the explanation is the fact, already mentioned, that low redshift objects have their distinctive spectral features in the bluer part of the spectrograph, a region where we experienced
more difficulties to assign redshifts. We further discuss the redshift dependence of the fraction of spectroscopically confirmed LBGs in section~\ref{sec:eff}.
The loss of efficiency at bright magnitudes led us to adopt a higher threshold, $r>22.7$ instead of $r>22.5$, for subsequent pilot surveys (see table~\ref{tab:cuts} and section~\ref{sec:extension}).

The location of LBGs with 
no robust redshifts in the color-color box of the target selection is given in the upper- left plot in figure~\ref{fig:contaminant_colourbox}.
The vast majority are close to the stellar locus (red dots), indicating 
probable star contamination. The location
of contaminants with spectroscopic type visually confirmed is presented in the three other plots of figure~\ref{fig:contaminant_colourbox}. ELGs  are preferentially found  at lower $g-r$ or lower $u-g$, while QSOs are more widespread in the box, with part of them at lower $g-r$, a region where no VI-typed LBG has been found. We also identify without any ambiguity a small number of stars which are very close to the stellar locus. In order to decrease the ELG and QSO contaminations in subsequent pilot surveys, we increased the $g-r$ color cut to  $g-r>0$ (see table~\ref{tab:cuts} and section~\ref{sec:refinedTS}).

We observed that contaminant ELGs are mostly low redshift objects ($0.05<z<0.15$) while contaminant QSOs are high redshift ones, $z>2-1.5$ (see figure~\ref{fig:VI_z}). Strictly speaking, these QSOs do not constitute a contaminant, as the DESI pipeline identifies them very efficiently and with precise redshift determination. Adding them to the main survey of DESI~\citep{QSO_TS_SV2023} would increase the density of QSOs at $z>2$ from $\sim 60$ per deg$^2$  to $\sim 100$ per deg$^2$. 

To complete this review of contaminants, out of 811 spectra visually inspected, it was not possible to assign a redshift to 115 (14$\%$) of them, 
either too faint or with no secure absorption line on a weak continuum. They were checked to have no specific location in the color-color selection box. 
These spectra 
were part of the test-sample used to check an improved dark current estimate for the spectroscopic pipeline (see section 4.2.1 in~\cite{Guy2023}). With the latter,
part of these spectra became exploitable.
The VI campaign on tile 80871 was not repeated with the 
improved dark current estimate but all results reported in the following of this paper benefit from that improvement.

\subsection{Composite rest-frame UV LBG spectra}
\label{sec:stacks}
For illustrative purpose, composite rest-frame UV spectra were built from the sample of visually inspected LBGs with a secure redshift ($Q\ge 2.5$). We used the VI campaign on the complete first TMG sample in table~\ref{tab:pilot} as well as  partial campaigns on the BXU sample. This results in a sample of 626 LBGs with a Ly$\alpha$ emission, whatever its strength, and 214 LBGs with no emission. Individual spectra were first blue-shifted to rest-frame using their VI redshift and then co-added (i.e. averaged) with an inverse-variance weighting. The result is presented in figure~\ref{fig:stacks}, separately for LBGs with and without Ly$\alpha$ emission. 

As mentioned in section~\ref{sec:vi}, VI redshifts rely on the main interstellar ionisation absorption lines, which explains that these absorption lines are aligned in the two composite spectra. 
As a result, the 
peak of the observed Ly$\alpha$ emission 
is displaced by about 2~\r{A} with respect to the Ly$\alpha$ wavelength of 1215.67~\r{A}. 
Two effects explain this displacement. First, a velocity offset is expected between the Ly$\alpha$ emission and the interstellar absorption lines, as a result of large-scale outflows in LBGs~\cite{Shapley2003}. Second, the  Ly$\alpha$ emission is partially absorbed in the host galaxy, which impacts the emission centroid. Altogether, 
at $z\sim3$, with a displacement of 2~\r{A}, redshifts based on absorption lines are expected to be lower by $\Delta z\sim 0.007$ ($dv \sim 500$~km s$^{-1}$)
than those based on the peak of the observed Ly$\alpha$ line.

\begin{figure} [t]
\centering
\includegraphics[width=1.0\textwidth]{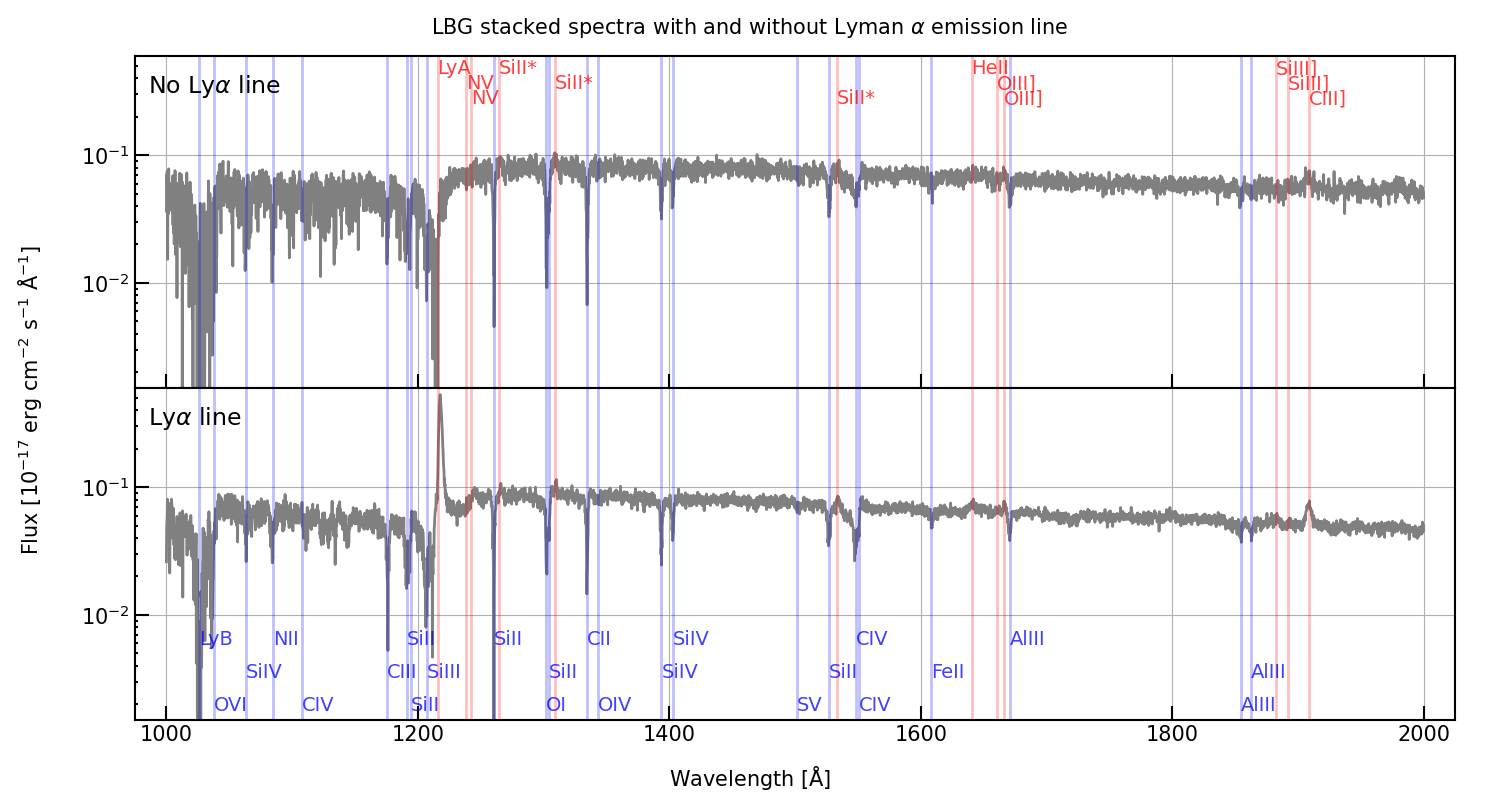} 
\caption{
Composite rest-frame UV spectra built from visually inspected LBG spectra.
{\it Top:} co-addition of 214 LBG spectra with no Ly$\alpha$ emission line.
{\it Bottom:} co-addition of 626 LBG spectra with a Ly$\alpha$ emission line.
The latter spectrum has been multiplied by $1.12$ such that the two composite spectra match between $1410$ and $1490$~\AA.
Blue (resp. red) labels indicate spectral lines in absorption (resp. emission). The bin size is 0.2~\r{A}.
}
\label{fig:stacks}
\end{figure} 



\section{Automated redshift measurement}
\label{sec:tools}

\subsection{Redshift and identification of LBGs with a Convolutional Neural Network}
\label{sec:CNN}

\subsubsection{Architecture of the CNN}

We developed  a convolutional neural network (CNN) to perform simultaneously a classification (LBG type or not) and regression task (determine the LBG redshift). This CNN is derived from the QuasarNET~\cite{QuasarNET18, 2020JCAP...11..015F} algorithm which is used for QSOs in DESI as part of the spectroscopic classification pipeline~\citep{Guy2023,RRQSO}.

The QuasarNet architecture  is inspired by 
those commonly used in Computer Vision, whose aim is to locate objects in images. As a first step, four convolution layers are used to extract spectral features.  Each of the four layers is separated by a Batch Normalization layer, which reduces the activations to a mean of zero and a standard deviation of one, thus improving the stability and convergence speed of the network. Following this Batch-Normalization layer, a non-linear Rectified Linear Unit (ReLu) activation function introduces the non-linearity required for the model to learn complex relationships between the data samples. Following these four convolution layers is a ``flatten'' layer, which transforms the output of the convolution layers into a one-dimensional tensor, which is then connected to a Dense layer with 100 neurons.  

The most interesting and distinctive part of QuasarNET is its ``line finder'', the final stage of the network. The idea is to create a doublet of two groups 
of neurons (``line detector'' and ``position finder'') for each absorption or emission line in the spectrum. 
Each group is made of $n_{boxes}$ neurons, each neuron corresponding to a slice in a region of the spectrum defined around the line.
In short, the ``line detector'' provides the confidence level of finding a line and the ``position finder'' provides the redshift corresponding to this line. 

For the application to LBGs, we used 14 absorption lines and two emission lines, 
as shown in table~\ref{tab:CNN_lines} (see also figure~\ref{fig:stacks}). Note that the wavelength used for the Ly$\alpha$ emission is about 2~\r{A} higher than that of the Ly$\alpha$ absorption to account for the redshift difference we measure when basing redshifts on absorption features.
QuasarNET was using $n_{boxes}=13$ for 7 spectral features. As we have 16 spectral features, we started with twice as many neurons and increased this number until the CNN efficiency and purity curve stopped improving, which led us to adopt $n_{boxes}=40$.

For each spectrum and each line, we keep the ``line detector'' neuron with the highest value and define this value as the confidence level ($CL$) for the line. We then rank the confidence levels for the 16 emission and absorption lines in decreasing order. 
A spectrum is declared as classified by the CNN as a LBG if the fifth confidence level 
is greater than a given threshold. 
The redshift associated to the line with the highest confidence level defines the LBG redshift.


\begin{table}[t]
\centering
\begin{tabular}{|c|c|}\hline
type & wavelengths (\r{A})\\ \hline
emission & Ly$\alpha$ $\lambda 1217.5$, \ion{C}{III} $\lambda 1909$ \\
absorption & Ly$\beta$ $\lambda 1026$, \ion{C}{III} $\lambda 1175$, \ion{Si}{II} $\lambda 1190$, \ion{Si}{III} $\lambda 1207$, Ly$\alpha$ $\lambda 1215.67$,
\\
 & \ion{Si}{II} $\lambda 1260$, \ion{O}{I} $\lambda 1302$+\ion{Si}{II} $\lambda 1304$, \ion{C}{II} $\lambda 1335$, \ion{Si}{IV} $\lambda\lambda 1394,1403$ \\
 & \ion{Si}{II}~$\lambda 1527$, \ion{C}{IV}~$\lambda 1548$, \ion{Al}{II} $\lambda 1670$\\
\hline     
\end{tabular}
\caption{Emission and absorption lines entering the training of the CNN used to confirm the type and determine the redshift of LBGs.
}
\label{tab:CNN_lines}
\end{table}

\subsubsection{Training of the CNN}
For QSO redshift determination with QuasarNET~\cite{QuasarNET18, 2020JCAP...11..015F}, the amount of data was not a problem, because the number of QSOs observed by SDSS was very large, with hundreds of thousands of spectra available. This is not the case for LBGs, which are much less widely observed. 
Visual inspection of spectra from the 2021 observation campaign gave a sample of around 1300 spectra confirmed as LBGs with high confidence ($Q\ge 2.5$) and around 200 contaminants, mostly QSOs and ELGs. 
To increase the sample size and enrich the sample in contaminants, we used a data augmentation method. In addition to the above confirmed LBG sample from VI, the final training sample also contains:
\begin{itemize}
\item Spectra of confirmed LBGs for different exposure times: spectra from individual exposures of the VI-confirmed LBGs were used to build co-added spectra for an effective exposure time of 2~hours. Those supplement the initial spectra which correspond to a 5~hour exposure time.
\item Shifted spectra of confirmed LBGs: as the visually inspected spectra essentially cover the redshift region from 2.3 to 3.3, we shifted these spectra to cover the entire region from 2.0 to 4.8. 
\item Spectra of ELGs and QSOs: we added spectra observed by DESI in the main survey, in redshift ranges derived from the visual inspection of section~\ref{sec:vi}, namely $0.05<z<0.15$ and $z>2$, for ELGs and QSOs, respectively (see figures ~\ref{fig:VI_z} and~\ref{fig:contaminant_colourbox}).
\item Simulated spectra of LBGs: composite spectra built from the co-addition of VI-confirmed LBG spectra 
were used to generate synthetic LBG spectra, with different Gaussian noise levels, over a continuous redshift distribution between 2 and 4.8.  
\end{itemize}
Regions where signals from two consecutive arms of the spectrographs are combined can exhibit spurious spikes in the measured flux and thus are not reliable enough for CNN training. In the above procedure, 40\r{A} wide regions covering these overlaps were thus masked in the spectra and replaced by  Gaussian noise.

With these additional data, the final training sample consists of more than 60,000 spectra. 
The CNN training is therefore carried out with this sample, but for a small sub-sample of visually inspected spectra and their repetitions in the augmented data sample. The sub-sample of visually inspected spectra is used to test the performance of the CNN. In this respect, our approach is similar to a k-fold cross validation technique. 

The removed sub-sample is the first sample of the TMG selection (see table~\ref{tab:pilot}), as the corresponding dataset (811 spectra) is the only complete sample that was visually inspected. This ensures that our estimates of the performance derived in the next section are unbiased. We split this sample into two sets, each containing half of the 426 VI-confirmed LBGs and half of the contaminants, to serve as two independent test-samples to measure the CNN performance. The full dataset is also used as a third training configuration. Each time, the sub-sample of VI-confirmed LBGs used as test-sample is removed from the CNN training.
Note that the architecture of the CNN is the same, whatever the training sample configuration.
The results are reported in figure~\ref{fig:eff_pur_vi}, which we discuss in the next section.

\subsubsection{Performance of the CNN}
\label{sec:CNN_perf}
We quantify the performance of the CNN by its \textit{efficiency} and \textit{purity}, which are
performance metrics  similar to parameters used in classification problems, the \textit{recall} and  the \textit{precision}, respectively.  
We define the \textit{efficiency}, $\varepsilon_{\rm CNN}$, as the fraction of VI-confirmed LBGs in the test-sample defined above that are selected by the CNN, and the \textit{purity}, $p_{\rm CNN}$, as the fraction of objects selected by the CNN in the test-sample that are VI-confirmed LBGs. 
Numerically, $\varepsilon_{\rm CNN}$ and $p_{\rm CNN}$ are defined as:

\begin{equation}
\label{eq:pur_eff}
\varepsilon_{\rm CNN} =  N^{LBG}_{\rm CNN}/N^{\rm LBG}_{\rm VI}
\quad \text{and} \quad
p_{\rm CNN}  =   N^{\rm LBG}_{\rm CNN}/N_{\rm CNN}
\end{equation}
where $N^{\rm LBG}_{\rm CNN}$ (resp. $N_{\rm CNN}$) is the number of VI-confirmed LBGs (resp. total number of objects) selected by the CNN in the test-sample. $N^{\rm LBG}_{\rm VI}$ is the number of VI-confirmed LBGs in the test-sample.

As mentioned in the previous section, we carried out three trainings of the CNN, based on the same CNN architecture and, in each case, we tested the CNN with a test-sample not used in the training. The left-hand plot in figure~\ref{fig:eff_pur_vi} shows the purity and efficiency curves obtained in the three cases, as a function of the confidence level threshold. As a reminder, we used as a test sample either one of the two halves of the TMG dataset for tile 80731 (set \#0 and set \#1 in the figure), or this complete dataset (set \#0 and \#1).
The right-hand plot in 
figure~\ref{fig:eff_pur_vi} is the efficiency-purity curve obtained when varying the confidence level threshold. We obtain a purity $\sim$85\% for a selection efficiency $\sim$70\%.

 \begin{figure} [t]
\centering
\includegraphics[width=1.0\textwidth]{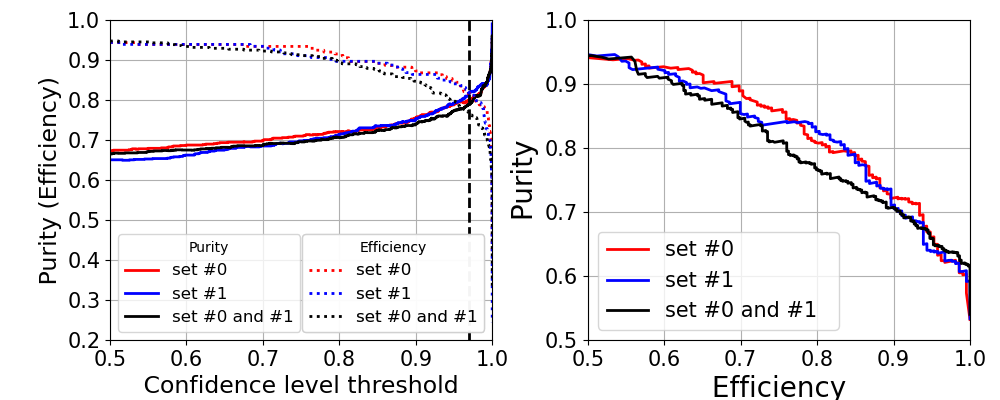} 
\caption{CNN selection purity and efficiency for LBGs as a function of the threshold on the confidence level provided by the CNN. A sample of around 60,000 confirmed LBG spectra (from both data and simulation) was used to train the CNN.  An independent sub-sample of spectra not used for the training is used to test the CNN performance. The three curves (red, blue and black) correspond 
to three configurations for the CNN training: without sub-sample \#0, without sub-sample \#1 and without both sub-samples \#0 and \#1, respectively. Each sub-sample \#0 or \#1 contains 213 visually inspected LBGs with good quality ($Q\ge 2.5$) as well as contaminants. The dashed vertical line indicates a confidence level of 0.97. The right-hand plot is the efficiency-purity curve obtained when the confidence level threshold is varied.
}
\label{fig:eff_pur_vi}
\end{figure}

In this paper, we use thresholds in the CNN confidence level mostly between 0.97 and 0.995 to achieve good selection efficiency while keeping contamination (by incorrect redshifts or objects different from LBGs) low. The final version of the CNN was established with the full LBG training sample described in the previous section.

\subsection{Template redshift fitting of LBGs with Redrock}
\label{sec:Redrock}

\begin{figure} [t]
\centering
\includegraphics[width=1.0\textwidth]{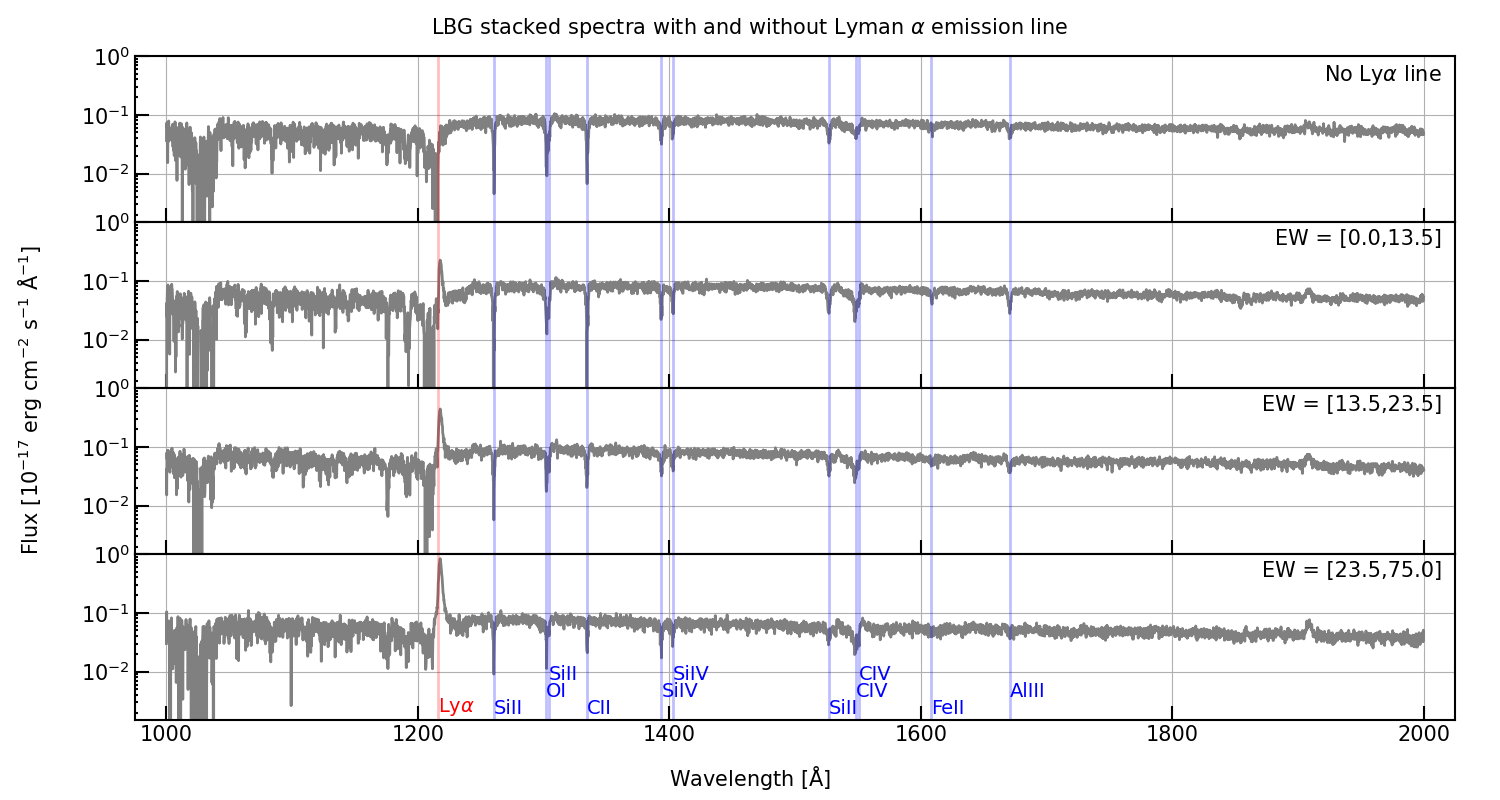} 
\caption{The four~rest-frame stacked spectra built from 840 visually inspected LBG spectra.
The first spectrum is built from LBGs with no~Ly$\alpha$ emission,
the last three from LBGs with Ly$\alpha$ emission in various EW intervals.
Blue lines show the main absorption lines and the red line shows the Ly$\alpha$ emission at 1215.67 \r{A}.} 
\label{fig:rr_stacks_forv20}
\end{figure}

\begin{figure} [t]
\centering
\includegraphics[width=1.0\textwidth]{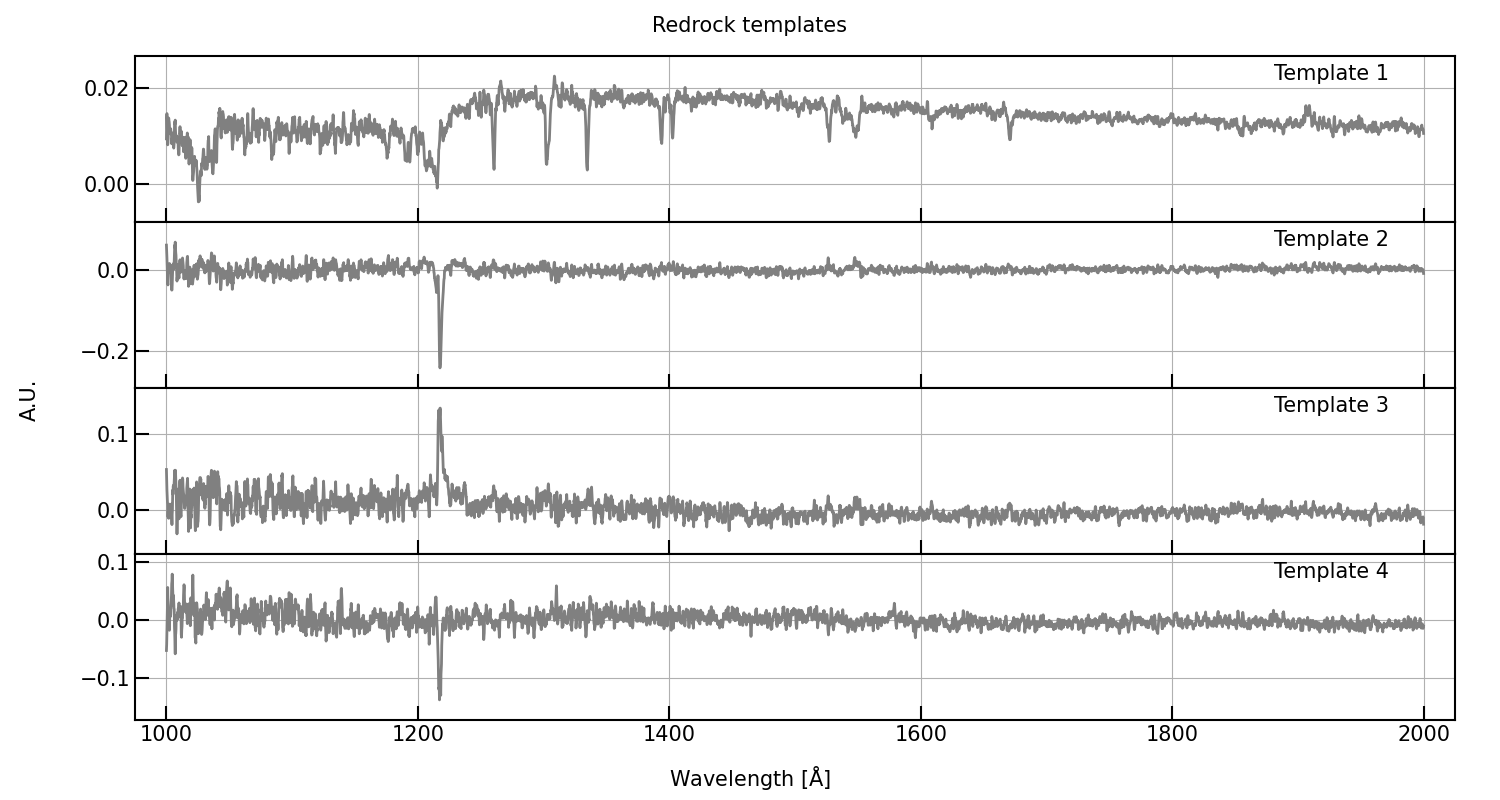} 
\caption{The four orthonormal templates used by Redrock for this analysis. The first template is the normalized stacked spectrum of LBGs with no~Ly$\alpha$ emission.}
\label{fig:rr_templates_v20}
\end{figure}

The precise LBG redshifts are obtained using the Redrock software\footnote{https://github.com/desihub/redrock} which is commonly used in DESI to perform spectral classification and spectroscopic redshift measurements for galaxies and quasars. 
A short description of the algorithm is provided in~\cite{Guy2023}. 
Redrock compares the measured spectra with a series of templates, performing for each of them a redshift scan followed by a reﬁned ﬁt to search for the best solutions. Current galaxy templates used by Redrock are based on stellar population synthesis and emission-line modeling of galaxies at $0<z<1.5$ and are not suitable for LBGs at redshifts above 2.

For this analysis, visually inspected LBG spectra are used to construct LBG-specific templates for Redrock, using the same sample of $840$~spectra ($626$~with an identified Ly$\alpha$ emission line and $214$ without) already mentioned in section~\ref{sec:stacks}. 
For LBGs with a Ly$\alpha$ emission, the rest-frame equivalent width (EW) of the line is first computed, to enable the sample to be divided according to EW values.
To do so, individual spectra are first blue-shifted to rest-frame using their VI redshift. The continuum on either side of the Ly$\alpha$ emission is estimated by linear fitting over the rest-frame wavelength ranges 1045-1170\r{A} and 1270-1500\r{A}, masking out the main absorption lines. Polynomial interpolation between the two linear fits provides the continuum estimate in the Ly$\alpha$ emission region. The fitted continuum is then subtracted from the spectrum and the profile of the emission 
fitted with an asymmetric Gaussian. The integrated flux in the fitted profile, 
normalized by the value of the continuum under the maximum of the fitted profile 
defines the rest-frame EW of the Ly$\alpha$ emission.

Four rest-frame stacked spectra are then produced as described in section~\ref{sec:stacks}, namely
one spectrum from the $214$~LBGs showing no emission, and three spectra from the $626$~LBGs with a Ly$\alpha$ emission, split into three equally populated EW intervals. 
  The stacked spectra, shown in~figure~\ref{fig:rr_stacks_forv20}, are filtered with an appropriate Savitzky-Golay filter~\cite{SavitzkyGolay} to reduce the noise.
A QR~orthonormalization was applied to the four spectra to obtain the Redrock templates (see figure~\ref{fig:rr_templates_v20}) such that the first template is the normalized stacked spectrum of the LBGs with no emission.
Redrock is run with these new templates, scanning a redshift range from~$2.0$ to~$4.8$ with a~$5 \times 10^{-4}$ logarithmic step, if no prior on the LBG redshift is known.

\subsection{Combined approach}
\label{sec:combi}

\begin{figure} [t]
\centering
\includegraphics[width=1.0\textwidth]{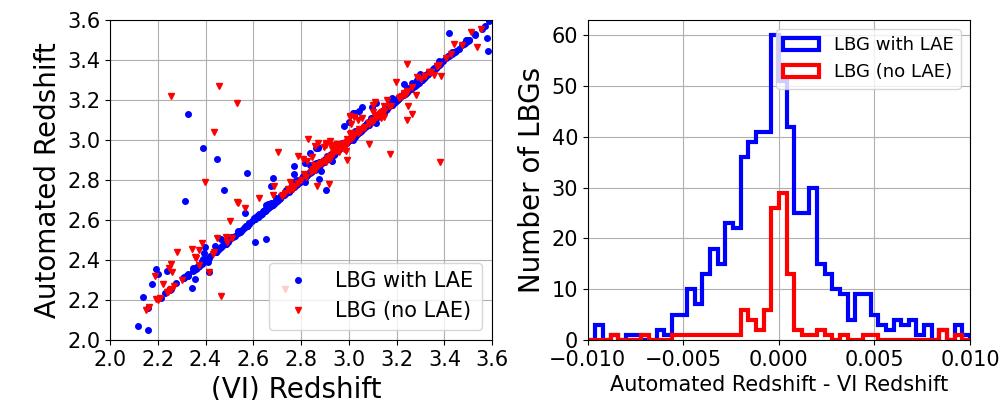} 
\caption{Comparison between VI redshifts and redshifts from the automated procedure. We used the set of visually confirmed LBGs of the TMG and BXU selections which have both good VI quality ($Q\ge 2.5$) and $CL>0.97$.
The blue (resp. red) dots and histograms correspond to LBGs with (resp. without) Ly$\alpha$ emission.}
\label{fig:reso_redshift_vi}
\end{figure}

LBG identification and redshift determination is a two-stage process. LBGs are first identified using the CNN described in section~\ref{sec:CNN}, which provides both a confidence level $CL$ and a redshift.
A flat prior of width $\pm 0.1$ around the CNN redshift is then fed on Redrock to obtain a precise redshift determination. Combining CNN priors and Redrock template fitting has already been used to build the DESI catalog of QSOs~\cite{QSO_TS_SV2023}, with very satisfactory results. 

Figure~\ref{fig:reso_redshift_vi} compares the redshifts obtained by the above combined automated procedure, $z_{\rm spec}$, with those obtained by visual inspection on the first observation campaign, $z_{\rm VI}$. In this comparison, VIed spectra are required to have both good VI quality ($Q\ge 2.5$) and $CL>0.97$.
The left-hand figure exhibits good agreement between the two redshift determinations, with outliers ($|z_{\rm spec}-z_{\rm VI}|/(1+z_{\rm VI})>0.01$) representing $\sim 12\%$ of the sample. 
The histogram of the difference between the two redshifts (see right-hand plot in figure~\ref{fig:reso_redshift_vi}) has a RMS of 200~km/s for LBGs with Ly$\alpha$ emission. 
Note that the automated procedure (CNN and Redrock steps) has been trained with 
the VI sample of the first observation campaign, so that these results can only be considered as a cross-check. We discuss redshift resolution and outliers further in section~\ref{sec:perfTS}.

Contamination by other spectral types has been checked on the whole VI sample, which contains 354 contaminants: 114 ELGs, 92 QSOs, 24 stars, 3 non-ELG galaxies and 121 featureless low signal-to-noise spectra. Selecting $CL>0.90$ reduces the contaminant number to 5, namely 4 ELGs and a weak continuum spectrum. A tighter selection $CL>0.97$ leaves only two contaminants, one ELG and the weak continuum spectrum. An efficient spectral typing is thus achieved by the CNN.

\begin{figure} [t]
\centering
\includegraphics[width=1.0\textwidth]{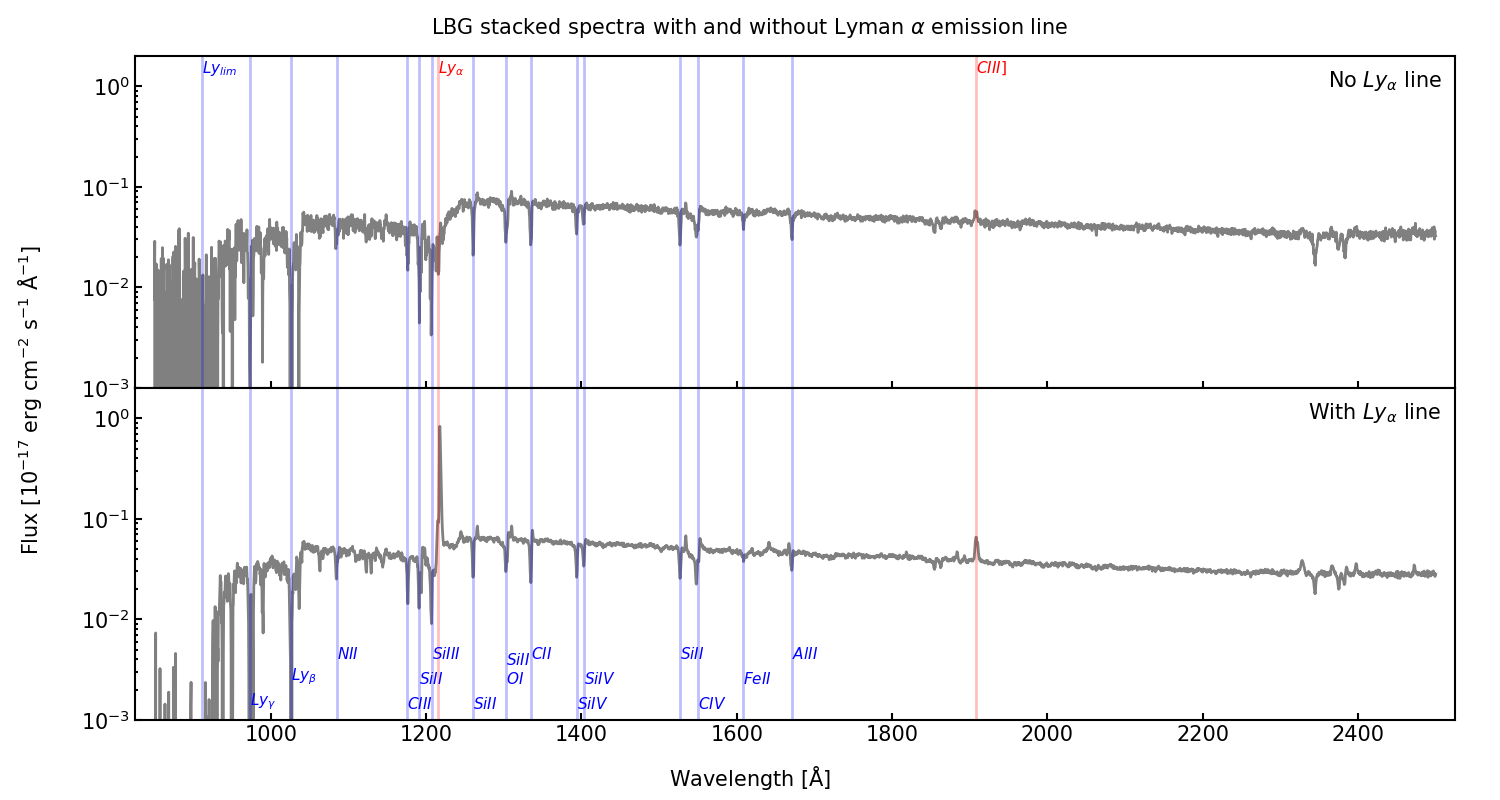} 
\caption{
Composite rest-frame UV spectra built from an enlarged sample of LBG spectra that include the VI sample as well as spectra selected by our automated selection with a cut $CL>0.99$.
{\it Top:} co-addition of $1145$ LBG spectra with no Ly$\alpha$ emission line.
{\it Bottom:} co-addition of $4848$ LBG spectra with a Ly$\alpha$ emission line.
Blue (resp. red) labels indicate spectral lines in absorption (resp. emission). The bin size is 0.4~\r{A}. The wavelength range extends from 850 to 2500~\r{A}. The Lyman limit (912~\r{A}) is clearly visible. 
}
\label{fig:stacks_enlarged}
\end{figure}

\begin{figure} [t]
\centering
\includegraphics[width=\textwidth]{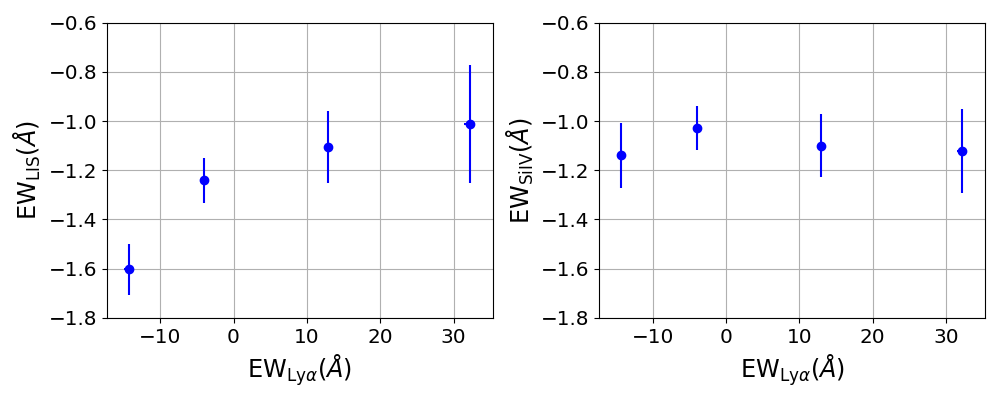} 
\caption{
Rest-frame absorption line equivalent width (EW) measurements as a function of the rest-frame Ly$\alpha$  EW, from visually inspected LBG spectra.
{\it Left:} average EW of low-ionization interstellar absorption lines.
{\it Right:} average EW  of the two high-ionization interstellar \ion{Si}{IV} absorption lines. 
}
\label{fig:ew}
\end{figure} 

\subsection{Further cross-checks of the automated procedure}
\label{sec:enlarged}
The combined approach previously described allows us to enlarge the data sample we can use to assess the performance of the $u$-dropout selection combined with spectroscopic type and redshift determination, as will be described in the next section.
To the VI sample used in section~\ref{sec:Redrock} (840 LBG spectra), we add all spectra selected by the automated procedure with a cut $CL>0.99$. This provides a total sample of 6495 spectra. 

Using Redrock-based redshifts for all spectra in this sample, we first blue-shift individual spectra to rest-frame. We apply criteria based on the Ly$\alpha$ emission profile fit to the rest-frame spectra to distinguish those with an emission line from those without. The corresponding stacks, constructed as in section~\ref{sec:stacks} are shown in figure~\ref{fig:stacks_enlarged}, to be compared with figure~\ref{fig:stacks}. There is a clear improvement in the signal-over-noise ratio of the stacks, allowing to enlarge their wavelength range so that the Lyman limit (912~\r{A}) is included. The Lyman break is clearly visible.

As a second cross-check, the rest-frame equivalent widths (EW) of the Ly$\alpha$ feature and the most visible absorption lines are measured on the individual spectra of the visually inspected sample, once blue-shifted to rest-frame using their Redrock redshift. 
To distinguish spectra with a Ly$\alpha$ emission from those without, we apply selections based on the Ly$\alpha$ emission profile fit and Ly$\alpha$ absorption measurement in the rest-frame spectra. Quality criteria are also required for all EW measurements (Ly$\alpha$ feature and absorption lines). 

For spectra with a Ly$\alpha$ emission line, the Ly$\alpha$ EW is measured as described in section~\ref{sec:Redrock} by means of a fit to the emission profile by an asymmetric Gaussian. This does not work as efficiently for spectra without an emission line. In that case, we rely on a more robust approach and make an integrated flux measurement below the continuum, starting at 1240~\r{A} and moving bluewards until the maximum of the absolute value of the integrated flux is reached.
The integrated flux is normalized by the continuum as measured in the range 1270-1500\r{A}, redward from the spectral feature. 

We also measure the EW of absorption lines redward from the Ly$\alpha$ feature, based on a fit with a Gaussian and a normalization by the continuum under the line. We measure the four low-ionization lines \ion{Si}{II} $\lambda 1260$, \ion{O}{I} $\lambda 1302$+\ion{Si}{II} $\lambda 1304$, \ion{C}{II} $\lambda 1335$ and the two \ion{Si}{IV} $\lambda\lambda 1394,1403$ high-ionization lines. For each spectrum, we thus have EW measurements for the Ly$\alpha$ spectral feature and for six absorption lines.

We then split the Ly$\alpha$ EW distribution into four quartiles. In each quartile, we compute the median values of the Ly$\alpha$ EW, of the four LIS absorption lines and the two \ion{Si}{IV} absorption lines.
Figure~\ref{fig:ew} presents the evolution of the median LIS and \ion{Si}{IV} EW measurements as a function of the median Ly$\alpha$ EW. The strength of the median LIS absorption clearly decreases as the Ly$\alpha$ EW increases, while the strength of the \ion{Si}{IV} absorption remains stable, a result in line with the literature~\cite{Shapley2003}.

Finally, an extra iteration was added to the combined approach of section~\ref{sec:combi} to use the enlarged sample of 6495 spectra to construct a new set of four stacks and four Redrock templates. We observe a slight improvement at the level of the redshift resolution and outliers in redshift, but no significant change in the clustering results reported in section~\ref{sec:clustering}. We thus decide to stick to the results based on the stacks and templates from the VI sample for the rest of the paper, and reserve the use of the templates from the enlarged sample for future work.

\section{\texorpdfstring{$u$}{u}-dropout target selections in second and third campaigns}
\label{sec:new_selections} 

\begin{figure} [t]
\centering
\includegraphics[width=\textwidth]{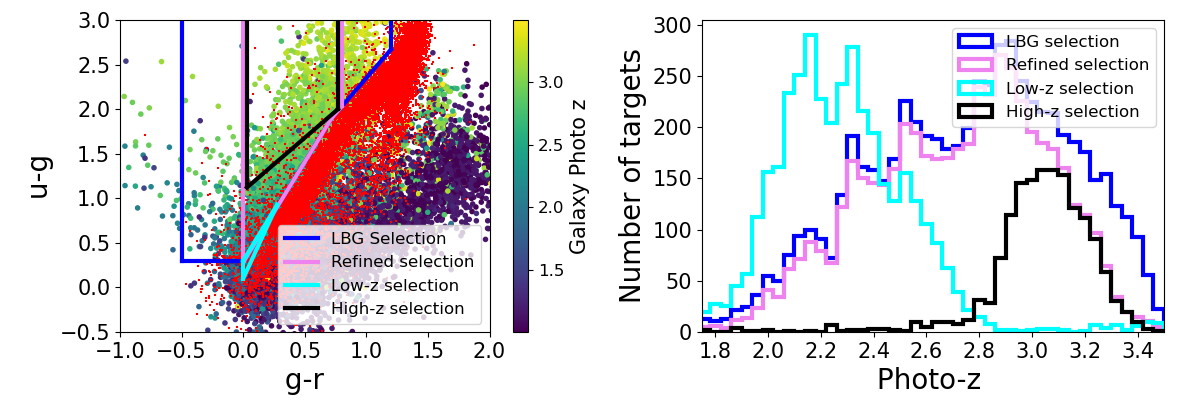}
\caption{{\it Left:} Color-color selection of LBG targets for the second DESI pilot survey on XMM-LSS (purple) compared to the initial selection used in the first pilot survey on COSMOS (blue). The cyan triangle corresponds to a low-$z$ extension. The black box represents a selection enriched in high redshift LBGs. Red dots are HSC stars and medium size dots are colored by photometric redshift~\cite{Arnouts99}. {\it Right:} Higher part of the \textsc{LePHARE} photometric redshift distribution for the first and second campaigns, in blue and purple, respectively. In cyan, photometric redshift profile of targets from the low-$z$ selection (triangle around $g-r\sim 0.1$ and $u-g \sim 0.4$). The black histogram corresponds to a subset of the standard LBG selection that enhances the fraction of LBGs with higher redshift.
}
\label{fig:ext}
\end{figure}

\subsection{Refined \texorpdfstring{$u$}{u}-dropout selection}
\label{sec:refinedTS}

Based on the analysis of the first campaign with visual inspection (see section~\ref{sec:res_vi}), we required $r>22.7$ and refined the color box definition of the $u$-dropout selection for the following campaigns (XMM-LSS field in winter 2022 and COSMOS field in spring 2023). 
The refinements applied to the XMM-LSS field (see details in table~\ref{tab:cuts}) are illustrated in figure~\ref{fig:ext} once converted to the $u$-filter, to allow direct comparison with the selections applied to the COSMOS field during the first campaign. The blue and purple polygons correspond to the color boxes for the first campaign and the subsequent campaigns, respectively. The main change consists in reducing the $g-r$ color cut from $-0.5<g-r<1.2$ to $0.0<g-r<0.8$. 

As shown in the right-hand plot in figure~\ref{fig:ext}, the photometric redshift distribution remains almost unchanged with the refined selection, but for LBGs with redshift around 3.4 which are removed, due to the higher cut-off on $g-r$. As the region $0.8<g-r<1.2$ is essentially populated by LBGs with no Ly$\alpha$ emission (see figure~\ref{fig:lbg_colourbox}), for which redshift determination is more difficult, overall the final loss of LBGs is expected to be low. 

The filters used for the $u$-band are not the same in CLAUDS imaging~\citep{CLAUDS19} for the XMM-LSS and COSMOS fields. The XMM-LSS field was observed with the old $u^\star$ filter, whereas both $u$ and $u^\star$ filters were used for the COSMOS field. To take this difference into account, the position of the inclined line in figure~\ref{fig:ext} has been slightly modified, but this is not visible in this figure. 

\subsection{Low-\texorpdfstring{$z$}{z} and high-\texorpdfstring{$z$}{z} selections}
\label{sec:extension}

As shown in figure~\ref{fig:redshifts}, the LBG redshift distribution peaks at $z\sim3.0$ and shows a strong decrease for $z<2.5$. 
In the second campaign, we tested an extension of the $u$-dropout selection to another color region in an attempt to increase the number of LBG targets at lower redshift i.e. $z<2.5$, a region of interest for DESI-II.

This 
low-$z$ selection represents a very small region of the $(g-r,u-g)$ color space and corresponds to the small cyan triangle centered at $g-r\sim 0.1$ and $u-g \sim 0.4$ in figure~\ref{fig:ext}. The right-hand plot in figure~\ref{fig:ext} presents the photo-$z$ distribution of these new targets, which covers the desired region in redshift. In addition, this figure shows that these new targets represent around one third of the total sample, illustrating the high density of targets in this region of color space. This new selection should offer a last advantage. As shown in figure~\ref{fig:lbg_colourbox}, this bluer region in $ g-r$ should contain more LBGs with a strong Ly$\alpha$ emission line, thus counterbalancing the intrinsic difficulty of measuring redshift for low-$z$ LBGs with DESI. 

Finally, also highlighted in figure~\ref{fig:ext} is a 
high-$z$ selection (color box in black) which gives a target sample in the photo-$z$ range between 2.8 and 3.4, a region where the redshift measurement efficiency is expected to be higher. 

In what follows, we assess the performance of the refined, low-$z$ extended $u$-dropout selection (pink and cyan polygons in figure~\ref{fig:lbg_extension}) in sections~\ref{sec:eff_def},~\ref{sec:eff} and~\ref{sec:eff2}. 
The performance of the low-$z$ selection alone and that of the high-$z$ selection (black polygon in figure~\ref{fig:lbg_extension}) are discussed in  section~\ref{sec:eff_ext}.
For the sake of simplicity, the refined, low-$z$ extended $u$-dropout selection is referred as to the extended $u$-dropout selection in the following.

\section{Target Selection Performance}
\label{sec:perfTS}
For this analysis, we use the automated method described in section~\ref{sec:tools} to confirm the LBG type and measure the redshift. We first define which criterion we use from the automated procedure to confirm a target as a true LBG and assess its performance with visual inspection. We then use that criterion to estimate the performance of the target selections applied during the second and third observation campaigns in the XMM-LSS and COSMOS fields, respectively
(see table~\ref{tab:pilot}). The performance studies are derived with the XMM-LSS data sample, which is larger than that collected on the COSMOS field. 

As an independent cross-check of our spectroscopic redshift measurements, we use the results from the COSMOS Lyman-Alpha Mapping And Tomography Observations (CLAMATO) survey~\cite{Horowitz22} which observed star-forming LBGs at $z\sim 2-3$ to create 3D maps of the IGM HI absorption using Ly$\alpha$ forest tomography. Some of their LBG targets were observed with DESI, allowing us to compare our automated redshifts to the CLAMATO ones.

\subsection{Spectroscopic confirmation definition and performance}
\label{sec:eff_def}
Throughout the section, we consider a target as a spectroscopically confirmed LBG when $CL>0.97$. 
For the 
extended $u$-dropout selection defined in section~\ref{sec:extension}, 
this retains $57\%$ of the targets. 
To quantify the LBG efficiency and purity of this spectroscopic selection, $\varepsilon_{\rm spec}$ and  $p_{\rm spec}$, a representative sub-sample of 300 spectra from the extended $u$-dropout selection was submitted to visual inspection. On this sample, the above criterion was found to keep $\varepsilon_{\rm spec}=83 \pm 3\%$ of the VI-confirmed LBGs ($Q\ge 2.5$) and to produce a sample where the fraction of such spectra is $p_{\rm spec}=90 \pm 2\%$. 

The fraction of redshift outliers was also studied with that VI sample by computing $|\Delta(z)|$ the difference in redshift between the VI and the automated procedure, divided by $(1+z_{\rm VI})$ where $z_{\rm VI}$ is the VI redshift. Note that this VI sample was not part of the training sample of the  automated procedure and thus constitutes an independent test-sample. The fraction of VI-confirmed LBGs with $|\Delta(z)|$ greater than 0.01 is $20\%$. This fraction becomes $10\%$ for VI-confirmed LBGs with Ly$\alpha$ emission.
Such a difference in redshift represents a velocity difference $dv=3000 \,\rm{km s^{-1}}$. This is the threshold considered in~\cite{2023AJ....165..124A,RRQSO} to set the requirement about catastrophic redshift failures for QSOs to be lower than $5\%$. This threshold in $dv$ is looser than that of other current DESI tracers due to the higher uncertainty in measuring redshifts from QSO broad emission lines. In the case of LBGs, the uncertainty is even higher since they are mostly faint targets with shallow absorptions and a Ly$\alpha$ emission line partly absorbed or absent. This explains why the fraction of redshift outliers for the same velocity threshold is much higher than for QSOs. 

In table~\ref{tab:vi_xmm} we explore different cuts in $CL$ and give the fraction of targets kept by the spectroscopic selection, $f_{\rm sel}$, as well as the spectroscopic selection efficiency, purity and fraction of outliers for both the selected spectra and those of them which have a Ly$\alpha$ emission. A $10\%$ fraction of outliers is obtained only for the highest threshold corresponding to a purity of $94\pm2\%$ and an efficiency of $56\pm 4\%$. In the reminder of this section, we keep the $CL$ threshold at 0.97 as it is not essential to maintain a low level of redshift outliers to estimate the target selection performance. 

\begin{table}[t]
    \centering
    \begin{tabular}{ccccccc}
    \hline
    $CL$ & $f_{\rm sel}$ & $\varepsilon_{\rm spec}$ & $p_{\rm spec}$  & {\small $f_{|\Delta z|>0.03}$} &  {\small $f_{|\Delta z|>0.01}$} & {\small $f^{\rm LAE}_{|\Delta z|>0.01}$} \\
   \hline 
   0.900 & 0.70 & 0.910 $\pm$ 0.022 & 0.803 $\pm$ 0.026 & 0.184 & 0.222 & 0.116 \\
   0.950 & 0.60 & 0.872 $\pm$ 0.026 & 0.890 $\pm$ 0.022 & 0.163 & 0.202 & 0.101 \\
   0.970 & 0.57 & 0.828 $\pm$ 0.030 & 0.899 $\pm$ 0.022 & 0.160 & 0.195 & 0.104 \\
   0.990 & 0.43 & 0.637 $\pm$ 0.038 & 0.909 $\pm$ 0.024 & 0.108 & 0.131 & 0.081 \\
   0.995 & 0.37 & 0.563 $\pm$ 0.039 & 0.935 $\pm$ 0.022 & 0.078 & 0.104 & 0.079 \\
   \hline
    \end{tabular}
    \caption{Results of visual inspection of a representative set of 300 spectra from the extended $u$-dropout selection in section~\ref{sec:extension}, observed in the XMM-LSS field. From left to right: threshold on the CNN Confidence Level used for spectroscopic confirmation, fraction of spectra selected with this CL threshold, purity and efficiency of the selected sample according to the definitions in eq.~\eqref{eq:pur_eff},
    fractions of redshift outliers with $|\Delta z|>0.03$, $|\Delta z|>0.01$ for good quality LBGs ($Q\ge 2.5$) and $|\Delta z|>0.01$ for good quality LBGs with a Ly$\alpha$ line. $\Delta z$ is defined as the ratio $(z_{\rm spec}-z_{\rm VI})/(1+z_{\rm VI})$ where $z_{\rm spec}$ is the redshift from our automated procedure and $z_{\rm VI}$ is the VI redshift. 
    }
    \label{tab:vi_xmm}
\end{table}

\subsection{Total efficiency of the extended \texorpdfstring{$u$}{u}-dropout selection}
\label{sec:eff}

To quantify the overall success of our ability to select 
and spectroscopically confirm LBGs, we hereafter define 
the {\it total efficiency} of the target selection as the ratio of the number of LBGs obtained by the automated method requiring $CL>0.97$, over the number of LBG targets. Note that the fraction of targets retained by the spectroscopic selection, $f_{\rm sel}$, introduced in the previous section and derived from the set of 300 VI-ed spectra, is a measure of the total efficiency, with moderate precision though, due to the small sample size. Thanks to the large sample size produced by the automated method applied to our complete set of collected spectra, the evolution of this total efficiency can be studied as a function of various LBG characteristics and spectroscopic observing conditions. In this section, we again focus on the extended $u$-dropout target selection defined in section~\ref{sec:extension}. 

Figure~\ref{fig:eff_z_teff} presents the total efficiency as a function of the effective exposure time of the spectroscopic observation and the target {\sc LePHARE} photometric redshift.
The effective exposure time is the amount of time it would take to reach a goal uncertainty in 'nominal' conditions for DESI, defined to be  a 1.1'' seeing, a sky background of 21.07 AB mag per square arc second in $r$-band, photometric conditions, observations at zenith, through zero Galactic dust reddening~\cite{Schlafly23}. In all figures showing variations in total efficiency as a function of photometric redshift, the redshift range is restricted to $2.1<z_{\rm phot}<3.5$. No restrictions are applied to the range of spectroscopic redshifts, as the CNN, which defines the spectroscopic redshift prior, is trained to find redshifts in the range between 2.0 and 4.8.

\begin{figure} [t]
\centering
\includegraphics[width=1.0\textwidth]{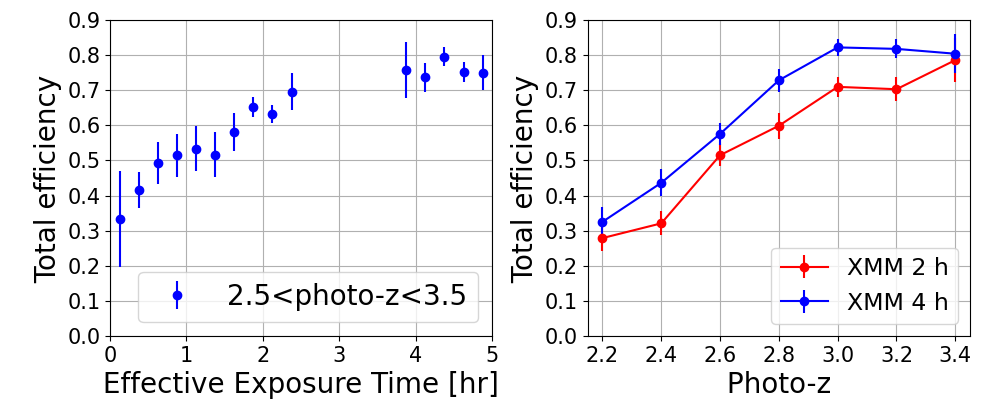} 
\caption{Total efficiency of the extended $u$-dropout target selection of section~\ref{sec:extension}, as a function of spectroscopic effective exposure time ({\it left}) and {\sc LePHARE} photometric redshift ({\it right}).}
\label{fig:eff_z_teff}
\end{figure}

The left-hand plot in figure~\ref{fig:eff_z_teff} shows that the total efficiency depends moderately on the effective exposure time beyond two hours of observation.
An effective exposure time of two hours therefore appears to be a good compromise between the total efficiency and the obtained density of spectra.
The right-hand plot in figure~\ref{fig:eff_z_teff} shows that, below $z\sim 3$, the lower the redshift, the lower the LBG detection efficiency. This is in line with the effect already observed with visual inspection (see figure~\ref{fig:fraction_lbg}), but confirmed here with a larger data sample, 
for different exposure times and for a different typing and redshift determination method. The same decrease of efficiency is also seen in the COSMOS field for the third observation campaign, which relied on the $U$-dropout selection with  no low-$z$ extension (see table~\ref{tab:cuts}). 

There are several possible causes for this behaviour: the limited accuracy of the photometric redshift may create migrations between redshift bins; the $u$-dropout target selection might be biased against low redshift emitters; the system throughput decreases in the blue channel of the instrument; the LBG population may evolve with redshift so that the fraction of Ly$\alpha$-emitting LBGs decreases at low redshift. The different causes are discussed hereafter.

\begin{figure} [t]
\centering
\begin{tabular}{cc}
\includegraphics[width=1.0\textwidth]{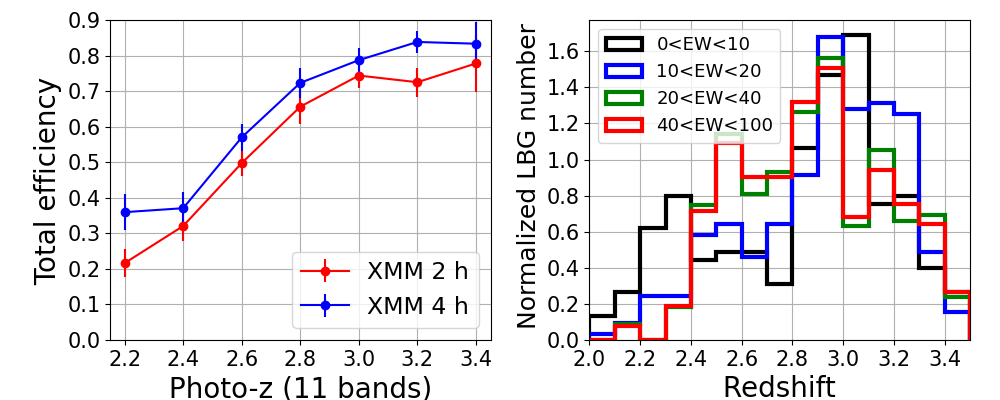} 
\end{tabular}
\caption{{\it Left:} Same as the right-hand plot in figure~\ref{fig:eff_z_teff} but restricted to the data sample with 11-band photometric redshifts. {\it Right:} Normalized spectroscopic redshift distributions of spectroscopically confirmed spectra of the extended $u$-dropout target selection of section~\ref{sec:extension}, divided in four bins of (positive) Ly$\alpha$ EW. 
}
\label{fig:eff_z11}
\end{figure}

\subsubsection{Photometric redshift accuracy}
The {\sc LePHARE} photometric redshifts for the CLAUDS objects in the XMM-LSS field use primarily 6 bands, the CLAUDS $U$-filter and the HSC-SSP $g,r,i,z,y$ bands~\cite{2019PASJ...71..114A}. But, for $\sim60\%$  of the XMM-LSS footprint, there are also photometric redshifts with 11 bands~\cite{2023A&A...670A..82D}, which use in addition the $Y, J, H, K_s$ bands from the VIDEO survey~\cite{2013MNRAS.428.1281J}. The analysis in this paper uses the 11-band photometric redshift when available, and the 6-band redshift otherwise. As the accuracy of the redshift improves with more bands, the efficiency study was repeated with the sub-sample of targets with 11-band redshifts. The result is shown in the left-hand plot of figure~\ref{fig:eff_z11}. The efficiency reduction with decreasing redshift below $z\sim3$ is still present: between redshifts 3.0 and 2.4, the 4-hour efficiency drops by $\sim$50$\%$, as in figure~\ref{fig:eff_z_teff}, although the efficiency curve shape is somewhat gentler. The effect of photometric redshift migration is thus sub-dominant. 

\subsubsection{Target selection bias due to band-pass crossing}
The $u$-dropout technique is likely to be more efficient when the $u$-dropout is due to the Lyman limit, that is for redshifts $z>2.5$ beyond which the Lyman limit enters the $u^*$ filter used for the XMM-LSS field. As the redshift increases beyond that value, the target selection efficiency is expected to increase as more of the absorbed region below the Lyman limit enters the $u^*$ filter. 
This impacts the pure number of targets that can be selected.

But this section deals with the total efficiency of the LBG selection, which is the fraction of the selected targets for which our automated procedure can assign a secure type and redshift. Even if less targets are selected, the ratio of those targets with correct type and redshift, over the number of selected targets, depends primarily on the spectroscopy step of the analysis and not on the variation of the dropout selection efficiency due to how much of the Lyman dropout region is in the appropriate photometric band. The only way the target selection can impact the total efficiency is the case when the target selection affects differently emitting and non-emitting LBGs, since it is more difficult to infer a type and redshift for the latter. This is unlikely to be the case for the target selection bias due to the Lyman break, which applies in the same way to both emitting and non-emitting LBGs.

On the other hand, the  
$u$-dropout technique may generate a bias against Ly$\alpha$-emitting LBGs at redshifts for which the Ly$\alpha$ feature is observed in the $u$-band, since Ly$\alpha$ emission would reduce the decrement in flux in the $u$-band w.r.t. the flux in the $g$-band. But this effect is limited to redshifts $z<2.4$, beyond which the Ly$\alpha$ feature is outside the XMM-LSS $u^*$-filter. 

This effect can be cross-checked with data on the COSMOS field. Indeed, the 2023 observation campaign on this field included a small sample of 232 low-redshift spectra, selected with a lower threshold in $U-g$ than our $U$-dropout selection for the same field: $U-g>2.2 (g-r) + 0.12$. The following requirements were also applied: $2.0 < z_{\rm phot} <2.5$ and $22.0<r<24.5$. 
The effective exposure time was the same for the two selections. 
The $U-g$ threshold for this additional sample is at least 0.2~mag lower than that used in our $U$-dropout selection (see last line in table~\ref{tab:cuts}). This difference encompasses the expected reduction in $u$-dropout due to the Ly$\alpha$ emission contribution, $\sim0.1$~mag~\cite{2023PASA...40...52F}. Repeating the efficiency study with this independent sample of spectra gives efficiencies in the redshift range $2.2<z<2.6$ compatible with those obtained for our $U$-dropout selection in the same field, and not higher ones as expected if there was a strong bias in the target selection due to the Ly$\alpha$ emission contribution.

\subsubsection{Total system throughput variations with wavelength}
\label{sec:throughput}
The total DESI system throughput has been reported to decrease below 4600~\r{A} (see Figure 26 in \cite{DESIOverview2022}), a region where the Ly$\alpha$ feature is expected to be observed at redshifts below 2.8. This decrease is dominated by fiber attenuation over a distance of 50~m, combined with the spectrograph throughput decrease below 4000~\r{A} (see Figure 17 in \cite{DESIOverview2022}), a region where the Ly$\alpha$ feature is expected to be observed at redshifts below 2.3.

We can use our efficiency measurements at effective times of 2 and 4 hours to check whether throughput variations could be the dominant factor of the efficiency drop below redshift 3. Assuming that the total efficiency behaves like the signal-over-noise ratio in the blue region of the spectrograph, variations in efficiency with exposure time may be either linear or square-root, if the noise is dominated by the detector or by the collected flux, respectively. Detector noise is not negligible in the blue part of the spectrograph, hence both hypotheses are worth considering. 
We would then expect a 30 to 50$\%$ drop in total efficiency by going from 4 to 2 hours, whereas the curves in figure~\ref{fig:eff_z_teff} show only a $\sim20\%$ drop over the redshift range $2.4<z<3.0$. It seems thus unlikely that the throughput variation explains all of the reduction observed in the total efficiency below redshift $z=3$. A simulation study would be necessary to obtain a more robust conclusion, but is beyond the scope of this paper.

\begin{figure} [t]
\centering
\includegraphics[width=\textwidth]{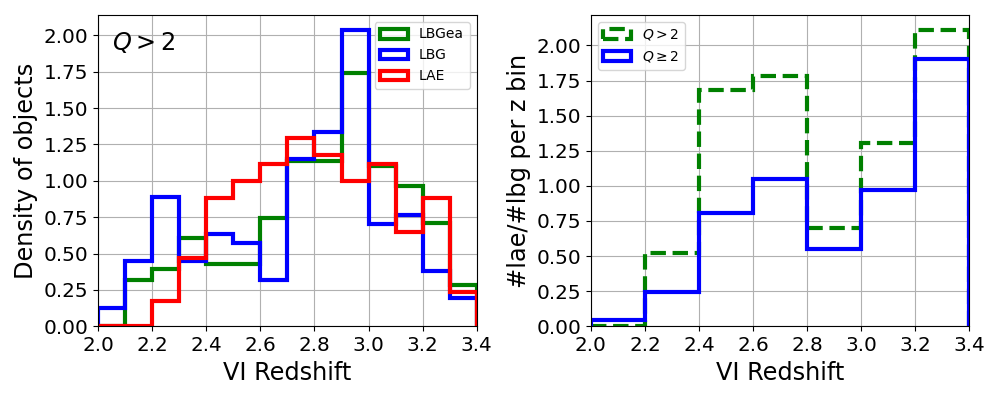}
\caption{{\it Left:} Normalized VI-redshift distributions of VI-confirmed spectra of the extended $u$-dropout target selection of section~\ref{sec:refinedTS}, split by LBG VI-type (LAE: Ly$\alpha$ emitter, LBG: non-emitting LBG, LBGea: low Ly$\alpha$ emission).  {\it Right:} Ratio of numbers of Ly$\alpha$ emitters and non-emitting LBGs as a function of VI-redshift, for good quality spectra (green) and including also spectra with unsecure resdhifts (blue).
}
\label{fig:xcheck_VI}
\end{figure}

\subsubsection{Population effect}
As a last possibility, there may be a possible population effect, as reported in the literature. Selecting LBG samples by their rest-frame UV colors at redshifts $z\sim2$ and $z\sim3$, was shown in~\cite{Reddy2008} to produce galaxy samples with a Ly$\alpha$ EW distribution representative of their parent galaxy samples, e.g. these color selections do not alter the intrinsic EW distribution of the parent galaxies. Using such samples and Ly$\alpha$ EW measurements, \cite{2023PASA...40...52F} showed that these LBG samples contain a significantly larger fraction of Ly$\alpha$ emitters at redshift $z\sim 3$ than at reshift $z\sim2$. Therefore, the drop in efficiency we observe below $z\sim3$ may be partly due to such an evolution effect, the reduced fraction of Ly$\alpha$ emitters with decreasing redshift leaving a larger fraction of LBGs with absorption, whose redshifts are more difficult to estimate.

As a cross-check of this hypothesis, the right-hand plot in figure~\ref{fig:eff_z11} shows the normalized spectroscopic redshift distributions of the spectra from the extended $u$-dropout target selection, 
split into four bins of Ly$\alpha$ EW, keeping aside those with negative EW for which the EW computation is not reliable. Objects with a large EW (strong emission) have a uniform redshift distribution between 2.4 and 3.2, while the distribution drops below $z<3$ for objects with a low EW (low or no emission). This shows that below $z<3$, type confirmation and redshift determination become more difficult for objects with no or low emission, while it remains at the same level for emitters. This, combined with a reduced fraction of emitters below $z\sim3$, due to population effect, could explain that the efficiency for the whole population drops so strongly below $z\sim3$. 

We note also that the distribution of emitters reduces to a negligible level below $z\sim2.4$. 
In the same redshift range, the distribution of non-emitters (or low ones) decreases more smoothly, because of higher noise in the blue channel, which makes the redshift determination less secure. Raising the CL threshold from 0.97 to 0.995, which increases the sample purity, we observe that the low-redshift part of the distribution of non-emitters (black curve) becomes smoother, suggesting that the peaks in the range $2.2<z<2.5$ are indeed due to incorrect redshift assignment.

Finally, we did a similar cross-check with the two complete VI samples analysed in this work (811 spectra from the TMG selection and 300 from the extended $u$-dropout selection on the XMM field). The left-hand plot in figure~\ref{fig:xcheck_VI} shows the object density with good VI quality, as a function of the VI redshift, separately for the three VI types defined in section~\ref{sec:vi}. The trends are similar to those in figure~\ref{fig:eff_z11}, which shows that these cannot be attributed to redshift determination or emitter definition.

The right-hand plot shows the ratio between the numbers of emitters and non-emitting LBGs, leaving aside the intermediate category whose typing is not as secure. We present this ratio for good VI-quality spectra only (green curve) and when we also include unsecure redshifts (blue curve). We remind that non secure VI-redshifts mean that their accuracy is higher than $\sim0.02$, which is ten times smaller than the redshift bin size in figure~\ref{fig:xcheck_VI}. With good quality spectra only, similar ratios are found at redshifts $\sim2.6$ and $\sim3.3$, showing no sign of a population effect. Adding non-secure redshifts increases the number of non-emitting LBGs by 66$\%$ and that of LBGs with low emission by 20$\%$, while the number of emitters increases by only 2$\%$. The right-hand plot in figure~\ref{fig:xcheck_VI} shows that these increases concern mainly the region $2.4<z<2.8$ and make the emitter fraction in this region lower than at $z\sim3.3$. This shows again that redshift determination is a significant issue in the $2.4<z<2.8$ range for non-emitting LBGs, and may indicate that there is in addition a population effect which disfavors emitters at redshift lower than 3.

\subsection{Total efficiency of the extended \texorpdfstring{$u$}{u}-dropout selection vs  magnitude and color}
\label{sec:eff2}

\begin{figure} [t]
\centering
\includegraphics[width=1.0\textwidth]{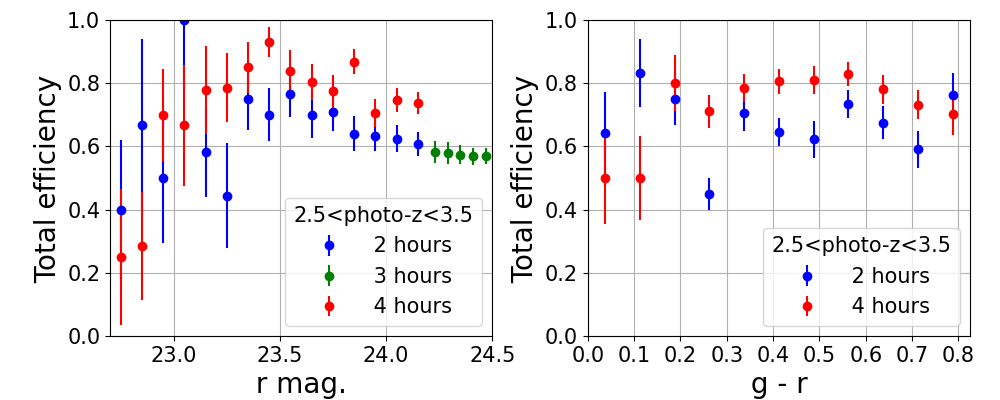} 
\caption{Total selection efficiency of the extended $u$-dropout target selection of section~\ref{sec:extension}, as a function of the $r$-band magnitude ({\it left}) and $g-r$ color ({\it right}) for a {\sc LePHARE} photometric redshift  between 2.5 and 3.5. 
In the left-hand plot, the green dots are from the extension of the $u$-dropout selection to fainter objects (see table~\ref{tab:cuts}).} 
\label{fig:eff_r_mag}
\end{figure}

The left-hand plot in figure~\ref{fig:eff_r_mag} shows that the total efficiency decreases as a function of the $r$ magnitude, as expected. However, for a reasonable exposure time of $\sim 2$ hours, the total efficiency remains approximately constant up to $r\sim 24.2$, allowing us to take this value as the optimal 
limiting magnitude for our observations. On the other hand, as shown in the right-hand plot in figure~\ref{fig:eff_r_mag}, the total efficiency exhibits no trend in $g-r$ color, 
which means that the LBG selection in $g-r$ cannot be further optimized.

Altogether, the target density of the extended $u$-dropout selection varies from 1100 to 2200 deg$^{-2}$ for a limiting magnitude from $r<24.2$ to $r<24.5$. For a cut in $r$ magnitude of 24.2, as advocated above, this target selection, followed by a spectroscopic confirmation with $CL>0.97$, would provide a density of $\sim 620$ deg$^{-2}$ LBGs 
in the range $2.3<z<3.4$ (see figure~\ref{fig:lbg_extension}).

\subsection{Performance of the low-\texorpdfstring{$z$}{z} and high-\texorpdfstring{$z$}{z} selections}
\label{sec:eff_ext}

\begin{figure} [tp]
\centering
\includegraphics[width=1.0\textwidth]{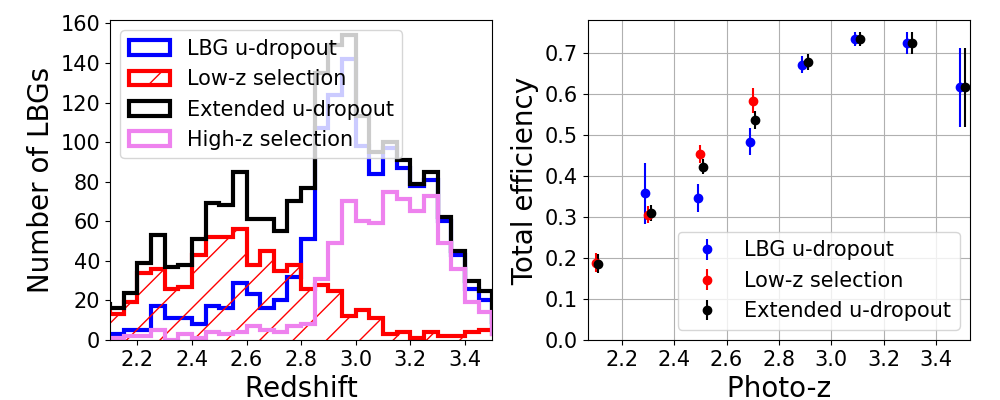} 
\caption{{\it Left:} Spectroscopic redshift distribution of LBG targets observed in the XMM-LSS field during the second observation campaign. The blue/red/black/violet histograms correspond to the $u$-dropout selection of section~\ref{sec:refinedTS}, the low-$z$ selection of section~\ref{sec:extension}, the combination of the two and the high-z selection of section~\ref{sec:extension}, respectively.
{\it Right:} Total efficiency as a function of the {\sc LePHARE} photometric redshift for the samples corresponding to the $u$-dropout selection, the low-$z$ selection and their combination, using the same color definition for the dots. For better visibility, a slight offset in redshift has been added to the data points for the various selections.
In both panels, data corresponding to 2 and 4 hours of effective exposure times have been used (see table~\ref{tab:pilot}).
}
\label{fig:lbg_extension}
\end{figure}

The results of the low-$z$ and high-$z$ selections introduced in section~\ref{sec:extension} are summarized in figure~\ref{fig:lbg_extension}, using again $CL>0.97$ as a spectroscopic confirmation. The left-hand plot shows the spectroscopic redshift distributions for the two selections compared with those for the refined $u$-dropout selection of section~\ref{sec:refinedTS} and the extended $u$-dropout selection of section~\ref{sec:extension}. 
As expected, the low-$z$ selection in the ($g-r=0.1$ and $u-g=0.4$) region allows us to obtain more LBGs with a redshift lower than 2.5. This is encouraging and demonstrates that we can tune the LBG redshift distribution from {\sc LePHARE}~\cite{Arnouts99} photometric redshifts. 

The right-hand plot in figure~\ref{fig:lbg_extension} shows that the dependence of the total efficiency as a function of the photometric redshift is essentially the same for the low-$z$ selection and for the refined  $u$-dropout selection. This result indicates that the LBG populations corresponding to the two selections are not different or that the differences are not sufficient to be reflected in the total efficiency.
As a result, the total efficiency for the low-$z$ selection is only around 35\% on average, to be compared with 70\% for the $u$-dropout selection at $z>2.8$. Such a low-$z$ selection, even if it selects the right redshift range, is not optimal in terms of efficiency.  

If the aim is not to increase the redshift coverage of the LBG sample, but to achieve the highest possible average total efficiency, a different strategy can be developed, as illustrated with the high-z selection shown in the left-hand plot in figure~\ref{fig:ext} (black polygon). There we move the lower side of the polygon upwards 
by applying a cut such as $u^*-g > 1.0*(g-r) + 1.8 $. This allows us to maintain 
a fraction of selected targets of $73\%$, an efficiency of $89\pm 4\%$ and a purity of $97\pm 2\%$, as reported in table~\ref{tab:vi_xmm_hiz}. The fraction of redshift outliers is also better, $13\%$ for the selected sample regardless of the LBG type and $4\%$ for LBGs with a Ly$\alpha$ emission.
With such a selection, the target density varies from 
325 to 650 deg$^{-2}$ for a limiting magnitude from $r<24.2$ to $r<24.5$. For a cut in $r$ magnitude of 24.5, this target selection, followed by a spectroscopic confirmation with $CL>0.97$ from our current automated algorithm, would provide a density of $\sim470$ deg$^{-2}$ LBGs 
in the range $2.8<z<3.4$. This would be sufficient for future programs such as DESI-II, or for a wider and deeper spectroscopic survey like Spec-S5~\cite{Schlegel2022} or WST~\cite{2023arXiv230816064B}.

\begin{table}[t]
    \centering
    \begin{tabular}{ccccccc}
    \hline
    $CL$ & $f_{\rm sel}$ & $\varepsilon_{\rm spec}$ & $p_{\rm spec}$  & {\small $f_{|\Delta z|>0.03}$} &  {\small $f_{|\Delta z|>0.01}$} & {\small $f^{\rm LAE}_{|\Delta z|>0.01}$}  \\
   \hline 
    0.900 & 0.87 & 0.972 $\pm$ 0.021 & 0.897 $\pm$ 0.035 & 0.116 & 0.159 & 0.066 \\
    0.950 & 0.77 & 0.930 $\pm$ 0.033 & 0.971 $\pm$ 0.020 & 0.091 & 0.136 & 0.035 \\
    0.970 & 0.73 & 0.888 $\pm$ 0.041 & 0.969 $\pm$ 0.021 & 0.095 & 0.127 & 0.036 \\
    0.990 & 0.62 & 0.761 $\pm$ 0.055 & 0.982 $\pm$ 0.018 & 0.056 & 0.093 & 0.040 \\
    0.995 & 0.57 & 0.719 $\pm$ 0.058 & 1.000 $\pm$ 0.019 & 0.020 & 0.059 & 0.041 \\
   \hline
    \end{tabular}
    \caption{Results of visual inspection of 89 spectra passing the high-z selection of section~\ref{sec:extension}, observed in the XMM-LSS field. From left to right: threshold on the CNN Confidence Level used for spectroscopic confirmation, fraction of spectra selected with this CL threshold, purity and efficiency of the selected sample according to the definitions in eq.~\eqref{eq:pur_eff},
    fractions of redshift outliers with $|\Delta z|>0.03$, $|\Delta z|>0.01$ for good quality LBGs ($Q\ge 2.5$) and $|\Delta z|>0.01$ for good quality LBGs with a Ly$\alpha$ line. $\Delta z$ is defined as the ratio $(z_{\rm spec}-z_{\rm VI})/(1+z_{\rm VI})$ where $z_{\rm spec}$ is the redshift from our automated procedure and $z_{\rm VI}$ is the VI redshift. 
    }
    \label{tab:vi_xmm_hiz}
\end{table}

\subsection{Sensitivity of the \texorpdfstring{$u$}{u}-dropout selection to \texorpdfstring{$u$}{u}-band depth }

\begin{figure} [t]
\centering
\includegraphics[width=1.0\textwidth]{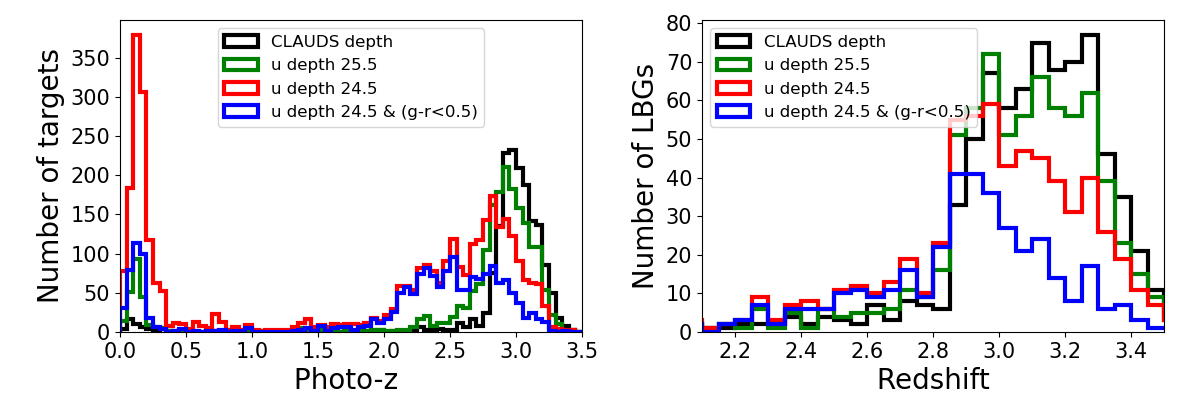} 
\caption{\textit{Left:} Photo-$z$ distribution of LBG targets selected by the high-$z$ selection defined in section~\ref{sec:extension}.
The total number of targets are 1685, 1996, 3655 and 1816 for the black, green, red and blue distributions, respectively, and the corresponding fractions of contaminants at $z_{\rm photo}<2$ are 5, 14, 41 and 26$\%$. 
\textit{Right:} Spectroscopic redshift distribution of LBGs selected by the high-$z$ selection and  spectroscopically confirmed by DESI. For both plots, the black, green, red histograms correspond respectively to the original CLAUDS depth, a 25.5 depth in $U$-band and a 24.5 depth in $U$-band.  For the latter configuration, an additional cut, $g-r<0.5$ is applied to obtain the blue histogram.}
\label{fig:high_z_depth}
\end{figure}

The high-$z$ selection is very robust in terms of both target selection and  redshift measurement by DESI. But this selection is based on CLAUDS imaging which provides deeper $u$-band imaging than will be available on large footprints at the start of future spectroscopic surveys,
DESI-II or Spec-S5. 
By the end of 2027, 
what will be available are CFIS~\cite{Ibata2017} data and likely one year of LSST data,
which both represent a $u$-band depth of $\sim 24.5$. 
By 2035, 
a depth of $\sim25.5$ can be expected with 10 years of LSST (LSST 10Y). 

To simulate the impact of such lower $u$-band depths on our LBG high-z $u$-dropout target selection, we degraded the CLAUDS imaging to depths of $24.5$ and $25.5$. 
Figure~\ref{fig:high_z_depth} shows the results in terms of both photometric redshifts for LBG targets and spectroscopic redshifts for confirmed LBGs. The left-hand plot exhibits a higher fraction of contaminants at $z\sim0.1$, namely $40\%$ (resp. $15\%$) for a $24.5$ (resp. $25.5$) $u$-band depth, respectively. The right-hand plot shows how the spectroscopic distribution of the actual high-z selected sample evolves when the $u$-depth of the corresponding targets is degraded. Although this plot misses the additional targets selected at $z_{\rm phot}<2.8$  with shallower depth (as seen on the left-hand plot), it confirms the loss of spectroscopically confirmed targets at redshifts above 2.8 already visible in the photometric redshift distribution on the left.

We note that the spectroscopic redshift distributions are quite similar for the $25.5$ $u$-band depth and the initial CLAUDS depth. This demonstrates the ability of selecting LBGs with a $u$-band depth corresponding to LSST 10, which is very encouraging for a project like Spec-S5.

Finally, as illustrated in the left-hand panel of figure~\ref{fig:high_z_depth}, 
for a $24.5$ $u$-band depth, an additional cut on $g-r<0.5$ (blue histogram) reduces the fraction of low-redshift ($z_{\rm photo}<2$) contaminants to $\sim25\%$, at the cost of a 
smaller mean redshift,
and provides $\sim10\%$ higher target density (over the whole redshift range) compared to the CLAUDS selection.

\subsection{Resolution on redshifts for the \texorpdfstring{$U$}{U}-dropout selection in COSMOS}
\label{sec:clamato}

\begin{figure} [t]
\centering
\includegraphics[width=1.0\textwidth]{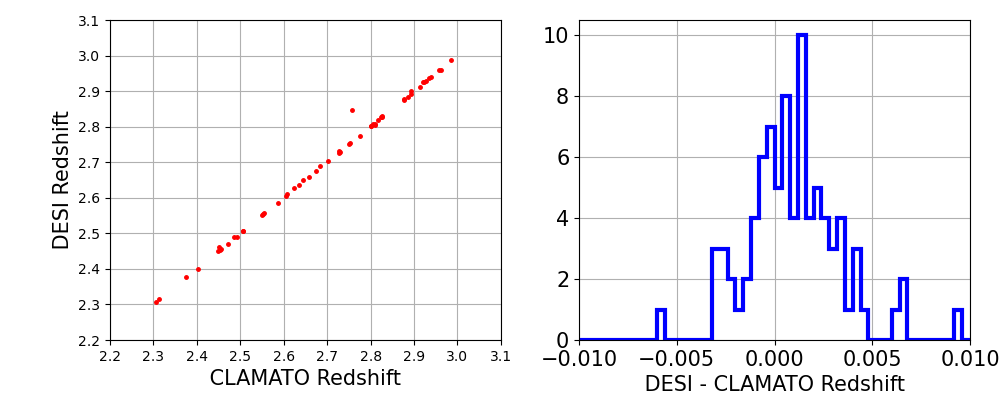} 
\caption{Comparison between  DESI automated redshifts  and CLAMATO redshifts on a sub-sample of 56 LBGs 
common to both surveys. The right-hand histogram is the difference between the DESI and CLAMATO redshifts. 
}
\label{fig:reso_redshift_clamato}
\end{figure}

DESI observations of the COSMOS field in 2023~(see table \ref{tab:pilot}) contain 105 LBGs observed (and confirmed as such) by CLAMATO~\cite{Horowitz22}, using the LRIS spectrograph on the Keck-I telescope at Maunakea. These observations were conducted in 2014-2020, for a total time allocation of 20.5 nights and 85~hours of on-sky integration. After reduction of the collected data, 600 spectra were extracted from the blue channel. Their redshifts and types were obtained by visual inspection and comparison with common line transitions and spectral templates (notably those of~\cite{Shapley2003}). Out of all these spectra, 359 were classified as galaxies with high confidence and secure redshifts above 2.

Even though the above 105 LBGs are not very bright ($r\sim 23.8$ on average) and have low redshifts ($\sim 2.6$ on average), they make a perfect control sample for testing our automated procedure. By applying the automated selection pipeline to the DESI spectra of these LBGs and requiring $CL>0.97$, we
retain 56 LBGs. The total efficiency for the CLAMATO LBGs, $55\%$, is a bit lower than what we found in section~\ref{sec:eff}, but these LBGs are fainter and at lower redshifts than our standard selection with CLAUDS.

Figure~\ref{fig:reso_redshift_clamato} shows a comparison between the redshifts measured by CLAMATO and with our automated procedure. Over 56 LBGs, there is only one outlier. The redshift resolution is 200~km/s with a small bias of $80\pm30$km/s, probably due to 
different conventions for defining redshifts.
These results are very encouraging and, as already shown in figure~\ref{fig:reso_redshift_vi}, confirm that our automated procedure is working correctly. 
This validation is even more well-founded, as this is a completely independent sample of spectra, never used in the CNN training nor in the production of our Redrock templates.

\subsection{Results of the \texorpdfstring{$g$}{g}-dropout selection}
\label{sec:g_eff}

\begin{figure} [t]
\centering
\includegraphics[width=1.0\textwidth]{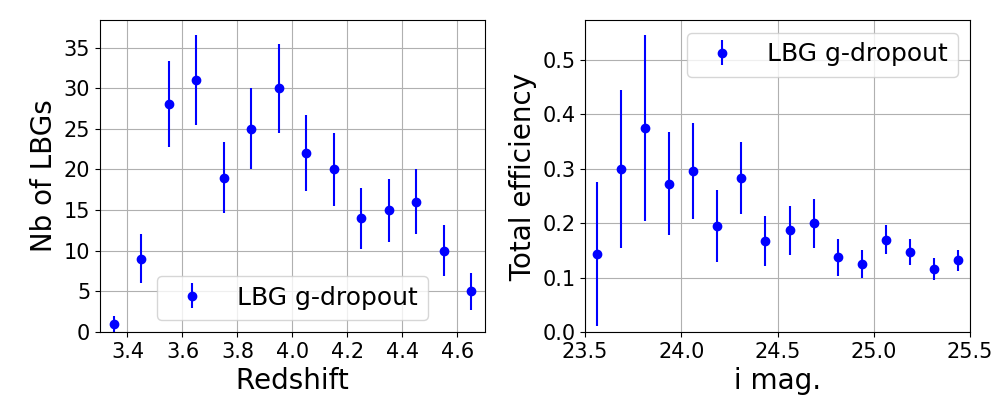} 
\caption{{\it Left:} Spectroscopic redshift distribution of LBG targets selected with the $g$-dropout selection of section~\ref{sec:g_dropout} and observed in the COSMOS field during the first campaign.  {\it Right:} Total efficiency as a function of $i$-band magnitude for the data sample with spectroscopic redshifts between 3.4 and 4.6.
}
\label{fig:res_g_dropout}
\end{figure}

In the first campaign, we observed a few hundreds targets with a $g$-dropout selection, described in section~\ref{sec:g_dropout}. The effective exposure time of around 5 hours was relatively short, given the low luminosity of these targets, $i\sim 25 $ on average. As a result, to analyze these spectra, we resort to a simpler and more robust approach than the combined automated method, since the CNN results are not optimal for this type of spectra. 
First, QSOs and low redshift $(z<0.8$) galaxies identified  by the standard DESI Redrock pipeline~\cite{Guy2023} are rejected. 
Then, we keep targets for which our version of Redrock, run with no CNN prior (see section~\ref{sec:Redrock}), returns a plausible redshift (\textsc{zwarning}=0) in the range $2.2<z<4.7$.
Finally, we require the significance of the detected Ly$\alpha$ line fitted by an asymmetric Gaussian profile to be greater than 3.

This way, we obtain a sample of 250 LBGs with a pronounced Ly$\alpha$ emission line (over 1612 targets). 
A visual inspection of this sample confirms that the purity is $>90\%$. Figure~\ref{fig:res_g_dropout} shows that the spectroscopic redshift distribution is centered at 
about four, as expected from the photometric redshift distribution in figure~\ref{fig:color_box_g_dropout}. The total efficiency, shown in the right-hand plot in figure~\ref{fig:res_g_dropout}, remains low, around 20\%, but this is to be expected given that these targets are very faint and that we adopted a very conservative, and therefore less efficient, approach to assess their type and redshift. In addition, the total efficiency 
slowly decreases at fainter $i$ magnitudes, confirming that the spectra 
become noisy for such magnitudes. 

Nevertheless, these results are very promising and demonstrate that LBGs can be selected in the desired redshift range,  $3.5<z<4.5$  with the $g$-dropout technique. This target selection would provide a low efficiency approach for DESI-II and therefore will be more appropriate for another generation of telescopes with a larger diameter primary mirror to fully exploit its potential.

\section{LBG clustering}
\label{sec:clustering}

The main objective of this paper is to promote LBGs as tracers of matter for future spectroscopic surveys in a redshift zone that is still relatively unexplored. 
However, the sample of spectroscopically confirmed LBGs collected during the DESI observation campaigns is large enough to allow a preliminary measurement of the linear bias of the selected LBGs to be derived from clustering measurements at relatively small scales.

\subsection{Data sample}
In the previous sections, the cut on the CNN confidence level to spectroscopically confirm a target as LBG was $CL>0.97$, ensuring a purity of 90\% (see table~\ref{tab:vi_xmm}) for the extended $u$-dropout selection. 
For this study, we select an even purer sample of LBGs by requiring $CL>0.995$, which increases the purity to almost 94\% and reduces the fraction of redshift outliers with $|\Delta z|>0.04$ from 20 to $10\%$.

This study is based on the two fields XMM-LSS and COSMOS, whose footprints are represented in figure~\ref{fig:footprint}.  The XMM-LSS field was observed during the second campaign and the COSMOS field during the first and third campaigns. LBG target selections were relatively similar in terms of colors, see table~\ref{tab:cuts}. 
During the fiber assignment process, LBG targets were considered mostly with high priority and thus are  moderately affected by fiber collisions with targets from other selections observed simultaneously.

In the remaining of this section, we restrict to the $u$-dropout selections, including the low-$z$ extension, but exclude the faint extension, sample 11 in table~\ref{tab:pilot}.
Among the corresponding 5973 (resp. 4510)
LBG targets which were observed by DESI in the COSMOS (resp. XMM-LSS) field, the spectroscopic selection previously mentioned retains 2209 (resp. 1405) LBGs, from samples 1 to 4 and 12 (resp. 7 to 10), as described in table~\ref{tab:pilot}. 

 \begin{figure} [ht]
\centering
\includegraphics[width=\textwidth]{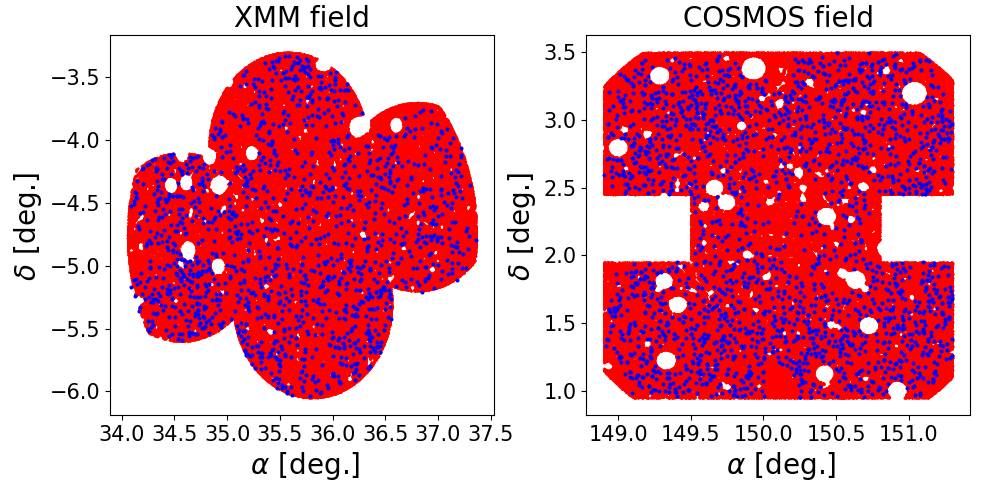}
\caption{Footprints used in the DESI analysis of the XMM-LSS and COSMOS fields of CLAUDS imaging. The red footprints correspond to the unmasked regions used in the 3D clustering analysis. The blue dots are the LBGs spectroscopically selected with $CL>0.995$.
}
\label{fig:footprint}
\end{figure}

\subsection{Correlation function}

The correlation function of LBGs 
is measured over a range of distances from a few Mpc$/h$ to a few tens of Mpc$/h$, which is not very sensitive to systematic effects related to CLAUDS imaging. Indeed, the XMM-LSS or COSMOS fields correspond to just four pointings, each pointing covering a 1.5 degree field-of-view. The observations for each pointing are very uniform, and it is only between pointings that 
imaging inhomogeneities may be observed. CLAUDS imaging is thus expected to present a very good homogeneity over the 
scales used to measure the correlation function. 

To take into account the CLAUDS imaging footprint, we generate a catalog of $5.10^{8}$ objects with ``random'' angular positions over the XMM-LSS and COSMOS fields, as observed by CLAUDS. As  can be seen in figure~\ref{fig:footprint}, we 
account for CLAUDS masks around bright stars. 
Each random object is assigned a redshift which is drawn from the measured redshift distribution $n(z)$ of the spectroscopically confirmed LBG sample obtained in the corresponding field.

We first compute the two point correlation function in two dimensions, $\xi(s,\mu)$, where $s$ is the galaxy pair separation and $\mu$ the cosine of the angle between the line-of-sight and the galaxy separation vector.  
The monopole of the two dimensional correlation function, $\xi_0(s)$, is then obtained by integrating $\xi(s,\mu)$ over $\mu$:
\begin{equation}
\xi_0(s) = \frac{1}{2}\int_{-1}^1 \xi(s,\mu) d\mu
\label{eq:LS_estimator}
\end{equation}
We rely on \textsc{pycorr}\footnote{\url{https://github.com/cosmodesi/pycorr}}, the DESI implementation of the~\textsc{corrfunc} package~\citep{Sinha2020} to compute the 
two dimensional correlation function and its monopole. We use 15 logarithmic bins in $s$ between 4 and 100~Mpc/$h$ and 200 linear bins in $\mu$ between -1 and 1.

Figure~\ref{fig:mocks_and_data} shows the LBG correlation function monopole 
for the two fields separately, with statistical uncertainties computed as described in the following section. 

\begin{figure} [t]
\centering
\includegraphics[width=0.45 \textwidth]
{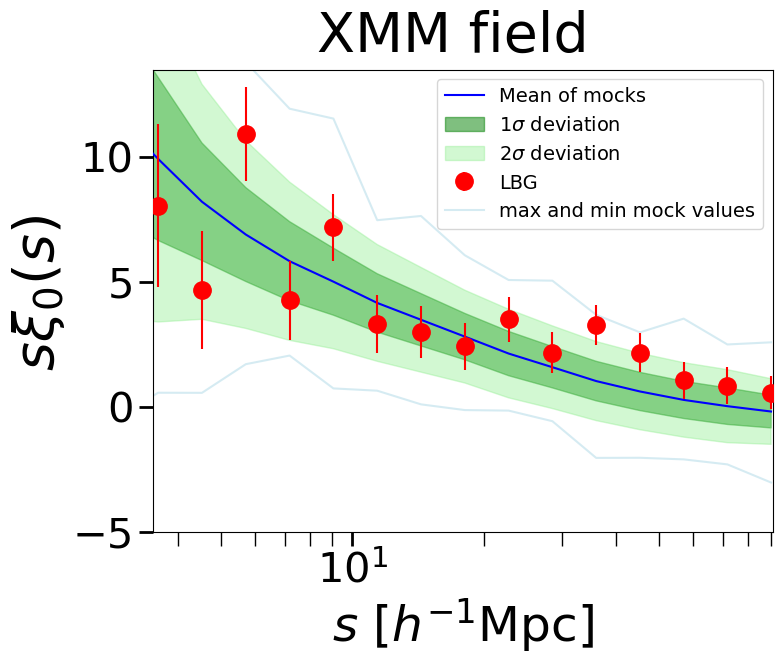}
\hspace{-0.23cm}
\includegraphics[width=0.45\textwidth]
{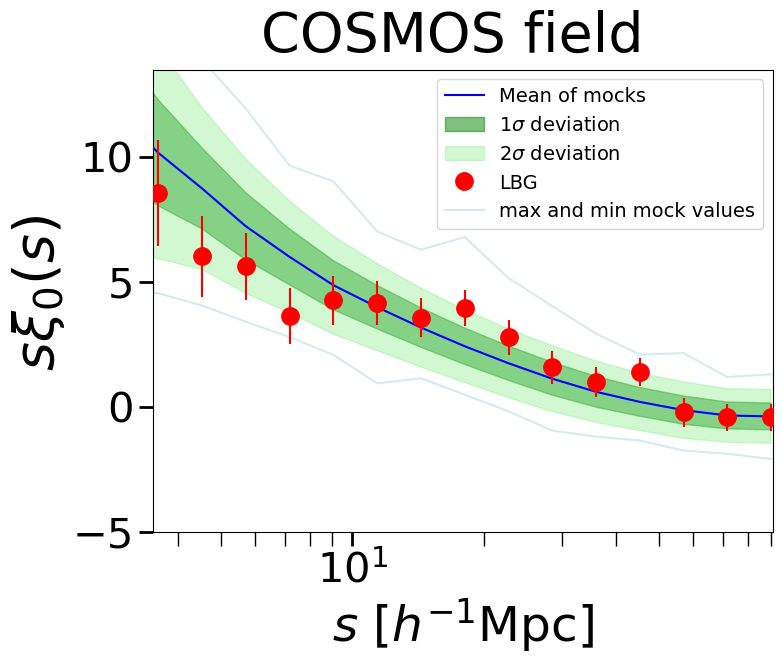}
\caption{LBG correlation function monopole, $\xi_0(s)$, measured in the XMM-LSS ({\it left}) and COSMOS ({\it right}) fields, using $CL>0.995$ as a spectroscopic selection (red dots). Light blue curves are the extremum values of the correlation function monopoles from mocks for each bin in s. The blue line is the mean of the mock monopoles. 
The green (resp. light-green) envelope is the 1$\sigma$ (resp. 2$\sigma$) dispersion of the mocks.
Error bars on red dots are statistical uncertainties deduced from the dispersion of the mocks.
}
\label{fig:mocks_and_data}
\end{figure}

\subsection{Statistical uncertainties}

We compute the covariance matrix using Lagrangian "lognormal" mocks, based on the \textsc{mockfactory} package in the \textsc{cosmodesi} environment. We generate mocks assuming a bias $b_{\rm LBG}=3.4$, a uniform redshift distribution in the range 
$2.3<z<3.5$ and a cubic box of size 800~(resp. 1000) Mpc$/h$ for the XMM-LSS (resp. COSMOS) field. 1000 mocks were created for each field. We checked with additional mocks in the XMM-LSS field that 1000 mocks were enough to obtain results independent of the number of mocks. 
 
We cut regions out of these two series of mocks to match the footprint and masks of their corresponding field. 
Then, in each series, we cut part of the mocks to have a redshift distribution similar to that of the LBG sample selected in the corresponding field.
To do so we infer a probability density function (PDF) 
from the redshift distribution $n(z)$ of the  
LBG sample selected in the field.
This PDF is then used to down-sample the redshift distribution of each mock, 
so that it matches that of data. We 
finally randomly down-sample the mocks to bring the number of 
objects close to that of the actual LBG data.

At this point, the monopole of the correlation function of each mock is computed as for data. 
For each field, the series of mock correlation function monopoles is then used to derive the covariance matrix for clustering measurements with the LBG sample in that field.
The corresponding correlation matrices obtained for XMM-LSS and COSMOS are presented in figure~\ref{fig:corr_matrix}. 
Clustering predictions from the mocks are also compared with data in figure~\ref{fig:mocks_and_data}, which illustrates the fact that the LBG correlation function monopole from data is well represented by the mocks we created. The statistical uncertainties of data in this figure are deduced from the dispersion of the mocks. When fitting the clustering from data in the next section, the $\chi^2$ computation uses the above covariance matrices corrected for the Hartlap effect~\citep{Hartlap07}.


\begin{figure} [ht]
\centering
\includegraphics[width=0.49\textwidth]
{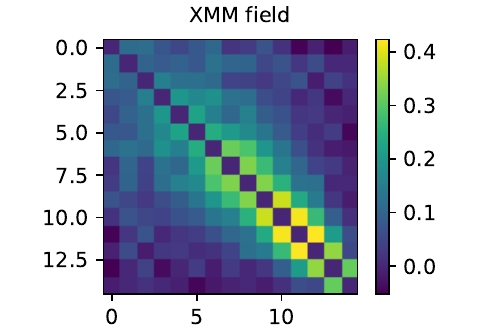}
\includegraphics[width=0.49\textwidth]
{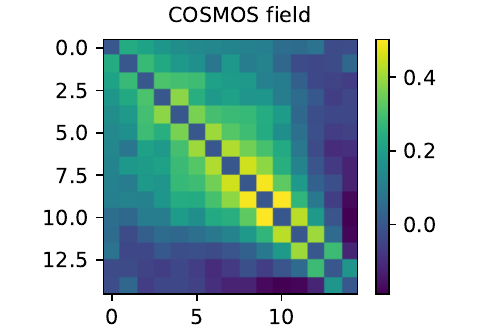}
\caption{Correlation matrices for clustering measurements, based on LBG mocks for the XMM-LSS ({\it left}) and COSMOS ({\it right}) fields. The matrix diagonals have been set to 0 by subtracting the identity matrix.}
\label{fig:corr_matrix}
\end{figure}

\subsection{Measurement of the LBG bias}

To measure the LBG bias, $b_{\rm LBG}$, we fit the measured $\xi_0(s)$ with a flat $\Lambda$CDM model, using the fiducial cosmology of~\cite{Planck2018}, corresponding to the following cosmological parameters: $h=0.6736$, $\Omega_m=0.3152$, $\Omega_b =0.04930$, $n_s=0.9649$. We use the DESI implementation of 
CLASS~\cite{lesgourgues2011} to generate a linear matter power spectrum, $P_{lin}(k)$ where $k$ is the norm of wavenumber ${\bf k}$. 
Linear redshift-space-distortions (RSD) can be accounted for through the Kaiser formula \cite{Kaiser1987} :

\begin{equation}
\label{Eq:Pmodel}
P_{\rm LBG}(k,\mu)=b_{\rm LBG}^2\; (1+\beta\mu^2)^2\;P_{lin}(k) \;,
\end{equation}
where $\mu$ is the cosine of the angle between ${\bf k}$ and the line of sight, $\beta = f/b_{\rm LBG}$, and $f \simeq \Omega_{m}^{0.55}(z)$ is the growth rate of structures in the $\Lambda$CDM model. 

A Fourier transform applied to eq.~\eqref{Eq:Pmodel} generates 
the two-dimensional correlation function $\xi(s,\mu)$ from which multipoles can be defined. As was shown in~\cite{Kaiser1987}, the monopole of the correlation function with linear RSD effects accounted for, $\xi_{\rm 0,LBG}(r)$ obeys the following equation:
\begin{equation}
\label{Eq:Xsimodel}
\xi_{\rm 0,LBG}(s) = b_{\rm LBG}^{2}(1 + \frac{2}{3} \beta + \frac{1}{5} \beta^{2}) \xi_{lin}(s),
\end{equation}
where $\xi_{lin}(s)$ is the Fourier transform  of $P_{lin}(k)$. We use the above formula to fit the measured clustering and derive values of $b_{\rm LBG}$. The method was first tested on mocks. On average over the 1000 mocks in each field, the input bias value is recovered with an offset (at most 0.07) which is below the expected uncertainty on the measurement. 

The results on data are shown in figure~\ref{fig:fit_bias}.
We measure $ b_{\rm LBG} = 3.48\pm 0.26 $ 
in the XMM-LSS field  and  $ b_{\rm LBG} = 3.09\pm0.24 $ in COSMOS, for a  mean redshift $z=2.9$ and a limiting magnitude of $r\sim24.2$.
The corresponding $\chi^2$ values are 23.9 for XMM-LSS  and 14.4 for COSMOS, both for 19 degrees of freedom, which correspond to p-values of 20$\%$ and 75$\%$, respectively.

The above results are in broad agreement with bias values measured in the literature for LBG samples. As an example, 
with a $u$-dropout selection similar to ours and a limiting magnitude of $r=24.5$, leading to a mean redshift $z\sim3$, \cite{Hildebrandt2009} found bias values in the range between 3.4 and 4.1, depending on the choice of the photometric redshift distribution used to de-project their angular correlation measurements. Other clustering results reported in the literature, for $u$-dropout selected LBG samples with $z\sim3$, span values from $2.22\pm0.16$~ for a combined Keck+VLT sample with $r<25.5$~\cite{2011MNRAS.414....2B}, to $3.5\pm0.3$ for the CFDF sample with $23.5<r<24.5$~\cite{2003A&A...409..835F}.
We note that the Keck+VLT sample has a lower bias value, which is likely to be due to their deeper sample limiting magnitude, as brighter LBGs are expected to be more clustered than fainter ones~\cite{2003A&A...409..835F}.


\begin{figure} [t]
\centering
\includegraphics[width=1.2\textwidth]
{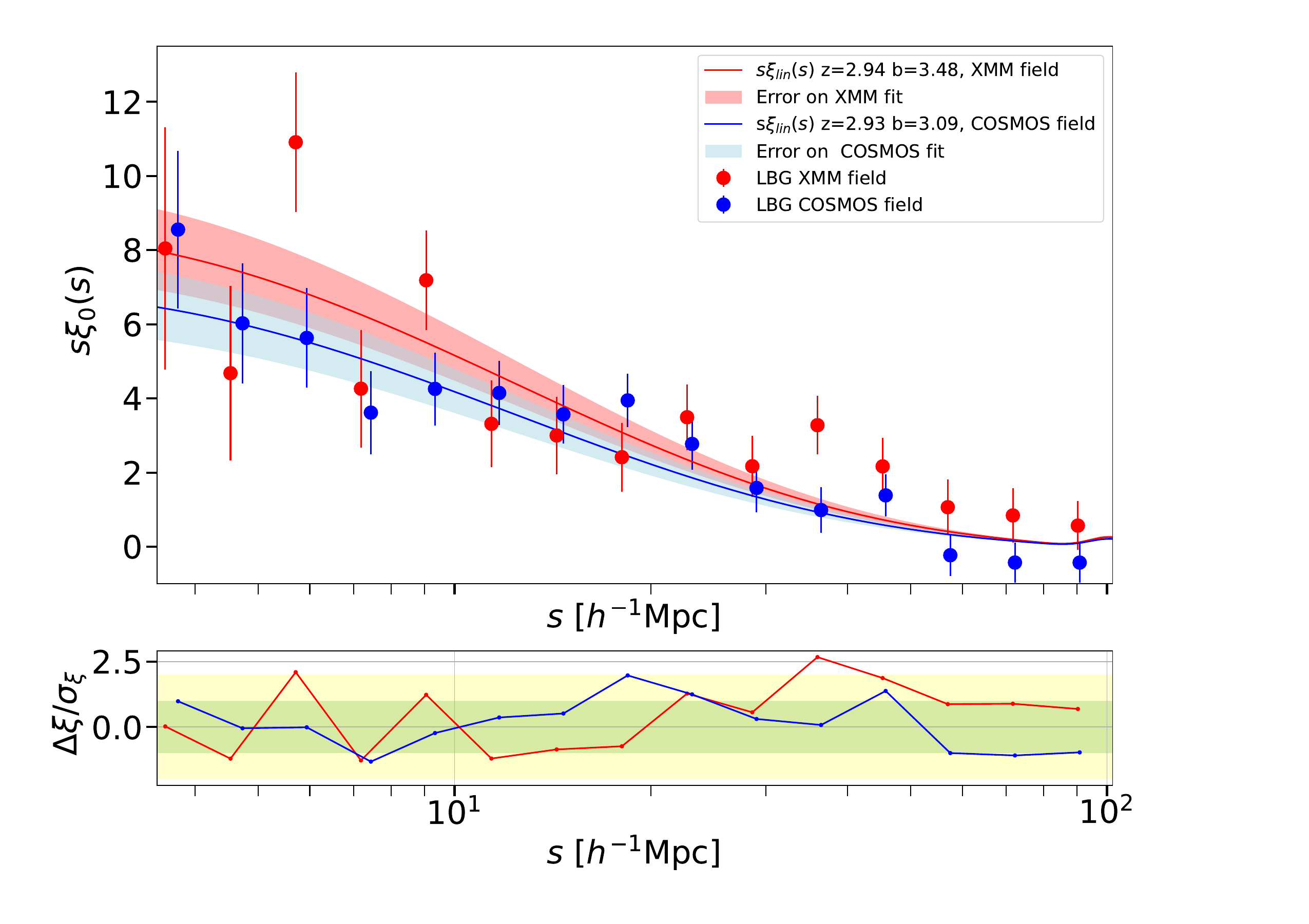}
\caption{{\it Top:} Monopole of the correlation function, $\xi_0(s)$, measured from the XMM-LSS (red) and COSMOS (blue) LBG data (dots with statistical errors). 
The solid lines are the results of the monopole fit including linear RSD effects from the Kaiser formula~\cite{Kaiser1987}. {\it Bottom:} Fit residuals normalized by data statistical uncertainties.
}
\label{fig:fit_bias}
\end{figure}

\begin{table}[h]
\centering
\begin{tabular}{|c|c|c|c|c|c|c|c|c|}\hline
$s$ [Mpc$/h$] & $4<s<90$  & $6<s<90$ & $7<s<90$ & $9<s<90$ & $11<s<90$ \\ \hline
XMM-LSS & 3.48$\pm$0.26 & 3.58$\pm$0.27 & 3.35$\pm$0.31 & 3.48$\pm$0.32 & 3.20$\pm$0.40 \\
COSMOS & 3.09$\pm$0.24 & 3.04$\pm$0.26 & 3.03$\pm$0.27 & 3.22$\pm$0.28 & 3.34$\pm$0.30 \\    
\hline     
\end{tabular}
\begin{tabular}{|c|c|c|c|c|c|c|c|}\hline
$s$ [Mpc$/h$] & $4<s<72$ & $4<s<57$ & $4<s<45$ & $4<s<36$ & $4<s<18$ \\ \hline
XMM-LSS & 3.48$\pm$0.26 & 3.47$\pm$0.26 & 3.47$\pm$0.26 & 3.47$\pm$0.26 & 3.41$\pm$0.27 \\
COSMOS & 3.10$\pm$0.24 & 3.12$\pm$0.24 & 3.16$\pm$0.24 & 3.13$\pm$0.25 & 3.12$\pm$0.25 \\    
\hline     
\end{tabular}
\caption{Measurement of the 
LBG bias for different fitting separation ranges, for the XMM-LSS and COSMOS data. The first column corresponds to the nominal measurement.}
\label{tab:cut_in_r}
\end{table}

\begin{table}[t]
\centering
\begin{tabular}{|c|c|c|c|c|c|c|c|c|}\hline
Confidence Level & 0.995 & 0.99 & 0.97 & 0.95 & 0.90  \\ \hline
XMM-LSS & 3.48$\pm$0.26 & 3.54$\pm$0.22 & 3.48$\pm$0.17 & 3.50$\pm$0.15 & 3.54$\pm$0.13 \\ \hline 
COSMOS & 3.09$\pm$0.24 & 3.02$\pm$0.22 & 3.55$\pm$0.16 & 3.63$\pm$0.15 & 3.68$\pm$0.13 \\
\hline     
\end{tabular}
\caption{Measurement of the 
LBG bias for different values of the 
threshold on the CNN Confidence Level ($CL$),  for XMM-LSS and COSMOS data. The higher the $CL$, the higher the LBG sample purity.  The first column corresponds to the nominal measurement.}
\label{tab:CL}
\end{table}

As a robustness test of our results, we varied the range in separations of the fit, to determine how sensitive the bias measurements are to the minimal and maximal bounds of the fitting range, as possible systematic effects in data could make the measurement unstable. 
As an example, if present, systematic effects from fiber collisions between LBG targets  would induce a dependence on the lower bound in $r$.  Similarly, if present, systematic effects from tiling would induce a dependence on the upper bound in $r$. Results are presented in table~\ref{tab:cut_in_r} and show 
that the bias measurements are consistent for most of the tested changes, showing good stability.

The dependence of the measurement on the purity of the LBG sample was also studied, by varying the 
minimal threshold applied to the CNN Confidence Level. Table~\ref{tab:CL} summarizes the results. 
We used the covariance matrices computed for the nominal case but rescaled by the squared ratio of the number of selected LBGs in the nominal case to that obtained with a different $CL$ threshold.
We observe that the $b_{\rm LBG}$ measurement remains stable for the XMM-LSS field, while for the COSMOS field 
the measured bias values vary significantly when the purity becomes lower. 
As a cross-check of this result, we repeated the clustering study using redshifts measured with the LBG templates based on the enlarged data sample, as described in section~\ref{sec:enlarged}. 
We used the same covariance matrices, as changes to the redshift distribution in both fields are negligible. We obtain $ b_{\rm LBG} = 3.39\pm 0.27 $ in the XMM-LSS field  and  $ b_{\rm LBG} = 3.06\pm0.25$ in COSMOS, in agreement with our nominal measurements. Stability tests with different $CL$ thresholds were repeated. The results for COSMOS show similar variations as those in table~\ref{tab:CL}, while for XMM-LSS results are no longer stable. The bias value is now observed to increase when purity degrades, as for the COSMOS field. 
Purity thus appears to have a significant impact on bias measurements.

As a conclusion from these stability tests, $4\, {\rm Mpc}/h <s<90 \,{\rm Mpc}/h$ and $CL>0.995$ appear to be optimal settings for the bias measurement from the LBG samples studied in this paper.

\section{Conclusion}
\label{sec:conclusion}

This paper introduced a target selection of high redshift LBGs based on deep broadband imaging from the CLAUDS and HSC-SSP surveys. In order to select LBGs at redshift above 2, we exploit the expected flux dropout shortward from the Lyman limit and the flux decrement shortward from the Ly$\alpha$ line, due to various absorption processes along the line of sight. We rely on a $u$-dropout technique and apply color cuts in the $u-g$ vs $g-r$ plane, as well as cuts in  $r$ magnitude. This selection and several extensions were tested in dedicated observation campaigns of the COSMOS and XMM-LSS fields with DESI, which collected a total of about 15,000 spectra. 

Around 13$\%$ of the spectra were visually inspected to characterize the target selection performance and set up an automated two-step algorithm that provides spectroscopic typing and redshift measurement for LBGs. 
The first step is a convolutional neural network (CNN) that was trained to identify sixteen characteristic LBG spectroscopic lines, among which the Ly$\alpha$ one  (either in emission or absorption),  
in order to assess the LBG type and provide a first estimate of the redshift. The latter is then used as a prior for template redshift fitting to provide a more precise LBG redshift measurement. 

To assess the target selection performance, we consider a target as a spectroscopically confirmed LBG when the CNN confidence level is greater than 0.97, which retains $57\%$ of the targets on average and provides an efficiency (resp. purity) of $83\%$ (resp. $90\%$) for the main $u$-dropout target selection, as measured from visually inspected spectra. With respect to visual inspection, the fraction of redshift outliers with 
$(z_{\rm spec}-z_{\rm VI})/(1+z_{\rm VI})$ greater than 0.01 ($dv=3000$km.s$^{-1}$) is $20\%$ overall, and $10\%$  for LBGs with Ly$\alpha$ emission. This reflects the difficulty to correctly measure the redshift of faint objects when only shallow absorption lines are present.

The main $u$-dropout selection provides LBG targets on a wide redshift interval, between 2.1 and 3.5. The selection efficiency was found to vary with redshift, 
raising from 25 to 80$\%$ between redshifts 2.1 and 3.0, and then remaining at this level up to a maximal redshift of 3.5, for an effective exposure time of 4 hours. The most likely reason for the drop in efficiency at redshifts below 3 is the difficult redshift determination in the range $2.1 < z < 3.0$ for non-emitting LBGs, combined with a possible evolution effect of the LBG population which would disadvantage emitters at redshifts lower than 3, as reported in the literature.
On the other hand, the efficiency does not depend strongly on the effective exposure time beyond two hours. Such an observing time thus seems a good compromise between total efficiency and the obtained density of spectra for LBG observations with the DESI instrument.

The 3d-clustering of the LBG sample collected in this paper was measured to estimate the linear bias of that sample, using a tighter CNN cut to restrict to higher purity, $94\%$. We found compatible values in the two fields, which average to $3.3 \pm 0.2 (stat.)$ for a mean redshift of 2.9 and a limiting magnitude in $r$ of 24.2. This is compatible with results from angular clustering found in the literature for similar $u$-dropout selections and limiting $r$ magnitudes, that cover a broad interval of bias values between 3.4 and 4.1.


As for prospects for a DESI-II phase, two different objectives can be envisioned. The main $u$-dropout selection described in this paper would deliver a high target density of $\sim$1100~deg$^{-2}$ in the redshift range $2.3<z<3.5$ for a limiting magnitude $r<24.2$. After spectroscopic confirmation, this would produce a LBG density $\sim$620 deg$^{-2}$. If the priority is put on efficiency rather than redshift coverage, the color cuts of this target selection can be easily modified by requiring a higher $u$-dropout at low $g-r$ color. This would produce a target density of $\sim$650~deg$^{-2}$ in the redshift range $2.8<z<3.5$ for a limiting magnitude $r<24.5$. 
Spectroscopic confirmation would select $73\%$ of these targets with $89\%$ efficiency and $97\%$ purity for LBGs, and give a fraction of redshift outliers of $13\%$ overall (resp. $4\%$ when a Ly$\alpha$ emission is present). This would provide a LBG density $\sim$470 deg$^{-2}$, an excellent basis for a high redshift clustering program.

\section*{Data availability}
All the material needed to reproduce the figures of this publication is available at this site: 
\url{https://doi.org/10.5281/zenodo.11387979}.

\acknowledgments
This material is based upon work supported by the U.S. Department of Energy (DOE), Office of Science, Office of High-Energy Physics, under Contract No. DE–AC02–05CH11231, and by the National Energy Research Scientific Computing Center, a DOE Office of Science User Facility under the same contract. Additional support for DESI was provided by the U.S. National Science Foundation (NSF), Division of Astronomical Sciences under Contract No. AST-0950945 to the NSF National Optical-Infrared Astronomy Research Laboratory; the Science and Technology Facilities Council of the United Kingdom; the Gordon and Betty Moore Foundation; the Heising-Simons Foundation; the French Alternative Energies and Atomic Energy Commission (CEA); the National Council of Humanities, Science and Technology of Mexico (CONAHCYT); the Ministry of Science and Innovation of Spain (MICINN), and by the DESI Member Institutions: \url{https://www.desi.lbl.gov/collaborating-institutions}. Any opinions, findings, and conclusions or recommendations expressed in this material are those of the author(s) and do not necessarily reflect the views of the U. S. National Science Foundation, the U. S. Department of Energy, or any of the listed funding agencies.

The authors are honored to be permitted to conduct scientific research on Iolkam Du’ag (Kitt Peak), a mountain with particular significance to the Tohono O’odham Nation.

This work also uses data obtained and processed as part of the CFHT Large Area U-band Deep Survey (CLAUDS), which is a collaboration between astronomers from Canada, France, and China. 
CLAUDS is based on observations obtained with MegaPrime/ MegaCam, a joint project of CFHT and CEA/Irfu, at the CFHT which is operated by the National Research Council (NRC) of Canada, the Institut National des Science de l’Univers of the Centre National de la Recherche Scientifique (CNRS) of France, and the University of Hawaii. CLAUDS uses data obtained in part through the Telescope Access Program (TAP), which has been funded by the National Astronomical Observatories, Chinese Academy of Sciences, and the Special Fund for Astronomy from the Ministry of Finance of China. CLAUDS uses data products from TERAPIX and the Canadian Astronomy Data Centre (CADC) and was carried out using resources from Compute Canada and Canadian Advanced Network For Astrophysical Research (CANFAR).

This paper is also based on data collected at the Subaru Telescope by the Hyper Suprime-Cam (HSC) collaboration and retrieved from the HSC data archive system, which is operated by the Subaru Telescope and Astronomy Data Center (ADC) at NAOJ. Their data analysis was in part carried out with the cooperation of Center for Computational Astrophysics (CfCA), NAOJ. 
The HSC collaboration includes the astronomical communities of Japan and Taiwan, and Princeton University. The HSC instrumentation and software were developed by the National Astronomical Observatory of Japan (NAOJ), the Kavli Institute for the Physics and Mathematics of the Universe (Kavli IPMU), the University of Tokyo, the High Energy Accelerator Research Organization (KEK), the Academia Sinica Institute for Astronomy and Astrophysics in Taiwan (ASIAA), and Princeton University. Funding was contributed by the FIRST program from the Japanese Cabinet Office, the Ministry of Education, Culture, Sports, Science and Technology (MEXT), the Japan Society for the Promotion of Science (JSPS), Japan Science and Technology Agency (JST), the Toray Science Foundation, NAOJ, Kavli IPMU, KEK, ASIAA, and Princeton University. 

The HSC collaboration members are honored and grateful for the opportunity of observing the Universe from Maunakea, which has the cultural, historical and natural significance in Hawaii. 


\bibliographystyle{JHEP}
\bibliography{references}{}

\begin{appendices}
\end{appendices}
\end{document}